\documentclass[12pt]{article}

\usepackage{amssymb,graphicx}

\usepackage{epsf}
\usepackage{graphicx,epsfig}
\usepackage{amsfonts}
\usepackage{amssymb}
\usepackage{cite}

\def\bc{{\bf c}}
\def\bk{{\bf k}}

\def\CA{{\cal A}}

\def\CL{{\cal L}}
\def\CO{{\cal O}}

\def\Ci{{\rm Ci}}
\def\Si{{\rm Si}}

\def\ty{{\tilde y}}

\def\Mpl{M_{pl}}

\def\half{\frac{1}{2}}

\makeatletter
\renewcommand\section{\@startsection {section}{1}{\z@}%
                                 {-3.5ex \@plus -1ex \@minus -.2ex}
                                   {2.3ex \@plus.2ex}%
                                   {\normalfont\large\bfseries}}
\renewcommand\subsection{\@startsection{subsection}{2}{\z@}%
                                   {-3.25ex\@plus -1ex \@minus -.2ex}%
                                     {1.5ex \@plus .2ex}%
                                     {\normalfont\bfseries}}
\renewcommand\subsubsection{\@startsection{subsubsection}{3}{\z@}%
                                   {-3.25ex\@plus -1ex \@minus -.2ex}%
                                     {1.5ex \@plus .2ex}%
                                     {\normalfont\itshape}}
\makeatother

\newcommand{\Letter}{
\setlength{\textwidth}{16.5cm}
   \setlength{\textheight}{22.6cm}
    \hoffset=-0.5in
\voffset=-2.1cm }

\Letter

\begin{document}
\newcommand{\be}{\begin{equation}}
\newcommand{\ee}{\end{equation}}
\newcommand{\bea}{\begin{eqnarray}}
\newcommand{\eea}{\end{eqnarray}}
\newcommand{\barr}{\begin{array}}
\newcommand{\earr}{\end{array}}

\thispagestyle{empty}
\begin{flushright}
\parbox[t]{1.5in}{
MAD-TH-06-3\\
UFIFT-HEP-06-9\\
SU-ITP-06/12\\
SLAC-PUB-11840\\
CU-TP-1147\\
hep-th/yymmnnn}
\end{flushright}

\vspace*{0.3in}

\begin{center}
{\large \bf Observational Signatures and Non-Gaussianities of  \\
\vspace{0.5cm}
General Single Field Inflation}

\vspace*{0.5in} {Xingang Chen$^{1,2}$, Min-xin Huang$^3$, Shamit
Kachru$^4$, and Gary Shiu$^3$}
\\[.3in]
{\em ${}^1$ Institute for Fundamental Theory\\
Department of Physics,
University of Florida,
Gainesville, FL 32611 \\[.1in]
${}^2$ ISCAP, Physics Department \\
Columbia University, New York, NY 10027, USA \\[.1in]
     ${}^3$ Department of Physics,
     University of Wisconsin \\
     Madison, WI 53706, USA\\[0.1in]
${}^4$ Department of Physics and SLAC, Stanford University\\
Stanford, CA 94305, USA } \\[0.3in]
\end{center}

\begin{center}
{\bf
Abstract}
\end{center}
\noindent
We perform a general study of primordial scalar non-Gaussianities in
single field inflationary models in Einstein gravity. 
We consider models
where the inflaton Lagrangian is an
arbitrary function of the scalar field and its first derivative, and
the sound speed is arbitrary.
We find that under reasonable assumptions,
the non-Gaussianity is completely determined by 5
parameters.  In special limits of the parameter space, one finds
distinctive
``shapes'' of the non-Gaussianity.
In models with a small sound speed, several of these shapes
would become potentially observable in the near future.
Different limits of our formulae
recover various previously known results.

\vfill

\newpage
\setcounter{page}{1}

\tableofcontents

\newpage

\section{Introduction}
\setcounter{equation}{0}

Inflation has been a very successful paradigm for understanding
otherwise puzzling aspects of big bang cosmology \cite{inflation,Lyth:1998xn}.
It can
naturally
solve the flatness, homogeneity and monopole
problems that otherwise seem to require a very high degree of fine tuning for
the initial state of our universe. Furthermore, inflation
generically predicts almost scale invariant Gaussian density
perturbations \cite{Mukhanov:1990me},
consistent with experimental observations of the
Cosmic Microwave Background. Future experiments can constrain and
distinguish between inflationary models
in several ways.

In the next few years, we can expect to see increasingly precise determination
of the scalar spectral index $n_s$ and its running\footnote{Already, there are strong
hints that models with a red spectrum are preferred \cite{WMAP}. Moreover,
if the spectral index runs from blue to red, then there should be 
an approximate 
coincidence \cite{Chung:2003iu}
between the length scale $k$ at which $n(k)-1=0$ and the length scale at which
the tensor to scalar ratio reaches a minimum. Such
a coincidence of scales, if observed, can 
put constraints on inflationary model building.}.
Planck will also lower bounds on the tensor-to-scalar ratio $r$ well into the
regime favored by models of large-field inflation \cite{Planck}.
However, $n_s$ and $r$
are just two numbers.  Although
their precise determination will be a tremendous
achievement, it will leave considerable ambiguity in reconstructing the
correct inflationary model.

In contrast, the non-Gaussian component of the scalar fluctuations is
characterized by a three-point function which is, a priori, a nontrivial
function
of three variables (momentum  magnitude and ratios) on the sky.
Furthermore, as demonstrated conclusively in \cite{Maldacena:2002vr,
Acquaviva:2002ud}, slow-roll models where the density perturbations
are produced by fluctuations of the inflaton itself,
predict negligible non-Gaussianity.
A detection of non-Gaussianity by the next generation of experiments
would therefore strongly favor either an exotic inflationary model,
or a model where density perturbations are generated by other dynamics
(as in curvaton \cite{curvaton} and modulated reheating
\cite{mr} scenarios).
A crude measure of non-Gaussianity is the number $f_{NL}$.  Values of
$\vert f_{NL} \vert \geq 5$ would almost certainly indicate some novelty in
the dynamics of the inflaton itself.

In this paper, we determine the most general non-Gaussian perturbations
possible in single-field inflationary theories.  We assume that the
inflaton itself
generates the density perturbations, and that the Lagrangian is a function
of the inflaton and its first derivative alone.
Under these assumptions, we prove that the full non-Gaussianity (at
the first
order in various slow-variation parameters)
is actually captured by five numbers.  These numbers characterize the
three-point function of fluctuations of the inflaton (or more precisely,
its gauge-invariant analogue).
For models with $c_s << 1$, which are known to produce the most significant
non-Gaussianities,
the result is stronger: the leading non-Gaussianity is characterized
by two numbers, and two different possible qualitative shapes in momentum
space; the subleading non-Gaussianity is characterized by three more
numbers.

The elegant, gauge invariant calculation of the non-Gaussianities for
slow-roll models appeared first in \cite{Maldacena:2002vr,Acquaviva:2002ud}.
The result is that slow-roll models produce a primordial $f_{NL}$
of ${\cal O}(10^{-2})$, too small to measure.
Creminelli stressed that models where the effects of higher derivatives 
are important may give larger $f_{NL}$ (under the assumption that the
non-Gaussianity is not diluted by the inflationary expansion itself, as 
happens in
some models \cite{CGS}).  However, this effect can only reach
$f_{NL} \sim {\cal O}(1)$ in the regime where effective field theory
applies \cite{Creminelli:2003iq}.  
A proof of principle that significantly larger $f_{NL}$ can
occur in sensible models
was provided by the work of Alishahiha, Silverstein and Tong
\cite{Alishahiha:2004eh}, who found that a fairly
concrete string construction \cite{Silverstein:2003hf} could yield
substantially larger $f_{NL} >> 1$.  
Different models in this class were constructed
\cite{Chen:2004gc,Chen:2005ad} in the context of 
warped compactification. The large non-Gaussianities in these models
are compatible with the current observational bound, but potentially
observable in future experiments
\cite{Alishahiha:2004eh,Chen:2005fe}.
(Another interesting model also providing
large $f_{NL}$, but so far resisting embedding in a UV complete theory,
appears in \cite{ghost}. This model does not belong to the general
class that we will study in Einstein gravity.)
Our results build on these papers and
further significant work by Seery and Lidsey
\cite{Seery:2005wm}, who found the fluctuation Lagrangian to cubic
order for a general class of Lagrangians, and of
Babich, Creminelli and Zaldarriaga \cite{Babich:2004gb}, who emphasized
the importance of analyzing the full shape of non-Gaussianities in $k$-space.

There are several motivations for completing such a general analysis.  Firstly,
it provides a null hypothesis against which to compare any future
measurement of the non-Gaussianity.  Secondly, several string-inspired
models, if realized in nature, will give rise
to a very characteristic measurable non-Gaussianity.
String theory is relevant here because models with significant non-Gaussianity
tend to be governed by higher-derivative terms, and a UV completion is
needed to make sense of such models (indeed, this is true even of
slow-roll models with negligible non-Gaussianities \cite{KKLMMT}, since
the slow-roll conditions are sensitive to Planck-suppressed corrections
to inflaton dynamics).
Our analysis should make it straightforward to work out predictions for
any such models.  When we wish to provide specific examples, we use
DBI inflation \cite{Silverstein:2003hf,Chen:2004gc} 
and K-inflation \cite{Kinflation}.

In addition, non-Gaussian fluctuations could contain a signature of any
departure of the inflaton from its standard Bunch-Davies vacuum. This
has been suggested as a possible signature of trans-Planckian physics,
and there has been much debate of the plausibility of such
modifications; some representative references are
\cite{Brandenberger,CGS,Shiu,Danielsson,Kempf,Kaloper,Burgess,BEFT,Porrati,CH,Mottola}.
It is a simple matter to translate any modification of the inflaton
wavefunction
into a modification of the three-point function, so the non-Gaussianities
could serve as a test of any proposed modification.

Finally, the structure of the three-point function can
in principle be determined by dS/CFT \cite{Andy}
or its generalization appropriate
to models with $c_s << 1$.  This could provide a useful laboratory
for studying holographic descriptions of dS space.
This perspective could be useful even if there is no exact relation
between the dS gravity theory and a dual field theory, since the useful aspects
of the duality for this purpose are purely kinematical.
Recent work in
this direction appears in \cite{Finn,Lidseyrecent}.  Our work on this
connection will eventually appear in a companion paper \cite{toappear}.

The organization of this paper is as follows.  In \S2, we present the
general class of Lagrangians that we will analyze (those that are an arbitrary
function of a single scalar field and its first derivative),  
and define the notation we will use in the rest of the paper.
While our analysis applies much more broadly,
in \S3 we describe three well-studied classes of inflationary theories
whose non-Gaussianities
we shall discuss in detail as special examples: 
slow-roll models, DBI models, and power-law
K-inflation models.
In \S4, we find the cubic fluctuation Lagrangian in appropriate gauge-invariant
variables for the most general single-field 
Lagrangian, and compute the non-Gaussianities.
Our main result is that there are only a few basic shapes, governed by
5 parameters in the most general model.
We evaluate our results
for the three special examples in \S5, and present the different qualitative
shapes of the non-Gaussianities that may occur.
In \S6, the effects of putting the inflaton in a vacuum other than the
Bunch-Davies vacuum are described.
We conclude in \S7.
The reader who is interested only in the class of Lagrangians studied and
the general structure of the non-Gaussianity for this class,
can confine her attention to sections \S2\ and \S4\ (which are more or less
self-contained).

\section{Inflation models with a general Lagrangian}
\label{SecSetup}
\setcounter{equation}{0}

To set up our notation, let us first review the formalism in
\cite{Garriga:1999vw} where a general Lagrangian for the inflaton field
is considered. The Lagrangian is of the general form
\begin{equation} \label{general}
S=\frac{1}{2}\int d^4x \sqrt{-g} \left[M_{pl}^2R +
2P(X,\phi)\right]~,
\end{equation}
where $\phi$ is the inflaton field and
$X=-\frac{1}{2}g^{\mu\nu}\partial_{\mu}\phi \partial_{\nu}\phi$.
The reduced Planck mass is $M_{pl}=(8\pi G)^{-\frac{1}{2}}$ and the
signature of the metric is $(-1,1,1,1)$.
The energy of the inflaton field is
\begin{equation}
E=2X P_{,X} -P ~,
\label{EPdef}
\end{equation}
where $P_{,X}$ denote the derivative with respect to $X$.
Suppose the universe is homogeneous with a
Friedmann-Robertson-Walker metric
\begin{equation}
ds^2=-dt^2+a^2(t)dx_3^2 ~.
\end{equation}
Here $a(t)$ is the scale factor and $H=\frac{\dot{a}}{a}$ is the
Hubble parameter of the universe. The equations of motion of the
gravitational dynamics are the Friedmann equation and the
continuity equation
\begin{eqnarray}
3M_{pl}^2H^2&=& E \label{eomH} ~,\\
\dot{E}&=&-3H(E+P) ~.
\end{eqnarray}
It is useful to define the ``speed of sound'' $c_s$ as 
\begin{eqnarray}
c_s^2 = \frac{dP}{dE}= \frac{P_{,X}}{P_{,X}+2X P_{,XX}}
\end{eqnarray}
and some ``slow variation parameters'' as
in standard slow roll inflation
\begin{eqnarray} \label{small}
\epsilon&=& -\frac{\dot{H}}{H^2}=\frac{X P_{,X}}{\Mpl^2 H^2}~, \nonumber \\
\eta &=& \frac{\dot{\epsilon}}{\epsilon H}~, \nonumber \\
s &=& \frac{\dot{c_s}}{c_s H} ~.
\end{eqnarray}
These parameters are more general than the usual slow roll
parameters (which are defined through properties of a flat potential,
assuming canonical kinetic terms), 
and in general
depend on derivative terms as well as the potential.
For example, in DBI
inflation the potential can be steep, and kinetically driven
inflation can occur even in absence of a potential. 
We also note that the
smallness of the parameters $\epsilon$, $\eta$, $s$ does not
imply that the rolling of inflaton is slow. 

The primordial power spectrum is derived for this general
Lagrangian in \cite{Garriga:1999vw}
\begin{equation} \label{power1}
P^{\zeta}_k=\frac{1}{36 \pi^2
M_{pl}^4}\frac{E^2}{c_s(P+E)}=\frac{1}{8 \pi^2
M_{pl}^2}\frac{H^2}{c_s\epsilon} ~,
\end{equation}
where the expression is evaluated at the time of horizon exit at
$c_s k=aH$. The spectral index is
\begin{equation} \label{index1}
n_s-1=\frac{d\ln P^{\zeta}_k}{d \ln k}= -2\epsilon-\eta-s ~.
\end{equation}
In order to have an almost scale invariant power spectrum, we need
to require the 3 parameters $\epsilon$, $\eta$, $s$ to be very
small, which we will denote simply as $\CO(\epsilon)$.
We note that in inflationary models with
standard kinetic terms the speed of sound is $c_s=1$, but here we do
not require $c_s$ to be close to $1$. For example, in the case of DBI
inflation, the speed of sound can be very small.
In the case of arbitrary $c_s$, the formula
(\ref{power1})(\ref{index1}) for the power spectrum and its index at
leading order is still valid as long as the variation of the sound
speed is slow, namely $s\ll 1$.
We will discuss this in more detail
in Sec.~\ref{SecQuad}.

The tensor perturbation spectrum $P^{h}_k$ and the tensor spectral index
$n_T$ are given by
\bea
P^h_k &\equiv& \frac{2}{3\pi^2} \frac{E}{\Mpl^4} ~,
\\
n_T &\equiv& \frac{d\ln P^h_k}{d\ln k} = -2 \epsilon ~,
\eea
and they satisfy a
generalized consistency relation
$\frac{P_k^h}{P_k^{\zeta}}=-8c_s n_T$. This is phenomenologically
different from standard inflation when the speed of sound is not
one.

\section{Several classes of models}
\setcounter{equation}{0}

In this section, we review three types of single field 
inflationary models. We
discuss the basic setups and results of the corresponding effective
field theories. These models will be used as primary examples after we
work out the general expression for
non-Gaussianities.

\subsection{Slow-roll inflation}
Slow-roll inflation models are the most popular models studied in
the literature. The effective action takes the canonical
non-relativistic form
\bea
P(X,\phi) = X-V(\phi) ~.
\eea
One achieves inflation by starting
the inflaton on top of a flat potential $V(\phi)$.
The flatness of this potential
is characterized by the slow roll parameters
\bea
\epsilon_V &=& \frac{\Mpl^2}{2} \left( \frac{V'}{V} \right)^2 ~,
\nonumber \\
\eta_V &=& \Mpl^2 \frac{V''}{V} ~,
\label{smallV}
\eea
which are required to be much less than one.
The energy
\bea
E=X+V \approx V
\eea
is dominated by the potential and the sound speed $c_s=1$.
During inflation the inflaton speed is determined by the attractor
solution
\bea
\dot \phi = - \frac{V'}{3H} ~.
\eea
This condition relates the slow roll parameters in (\ref{smallV}) to
the slow variation parameters in (\ref{small}),
\bea
\epsilon = \epsilon_V ~,~~~~ \eta = -2 \eta_V + 4 \epsilon_V ~.
\label{smallrelation}
\eea
The primordial scalar and gravitational wave power spectrum are both
determined by the potential
\bea
P^\zeta_k &=& \frac{1}{12\pi^2 \Mpl^6} \frac{V^3}{V'^2} ~, \\
P^h_k &=& \frac{2V}{3\pi^2 \Mpl^4} ~.
\eea
The spectral indices and the running can be
computed using the relation
\begin{equation}
d \ln k=H dt
=\frac{H}{\dot{\phi}}d\phi ~,
\label{lnkphi}
\end{equation}
and we get
\bea
n_s-1 &=& \frac{d\ln P^\zeta_k}{d\ln k}
=\Mpl^2 \left( -3\frac{V'^2}{V^2} + 2 \frac{V''}{V} \right)
~, \nonumber \\
\frac{d n_s}{d\ln k} &=& \Mpl^4 \left( -6 \frac{V'^4}{V^4} + 8
\frac{V'^2 V''}{V^3} -2 \frac{V' V'''}{V^2} \right) ~, \nonumber \\
n_T &=& \frac{d\ln P^h_k}{d\ln k} = -\Mpl^2 \frac{V'^2}{V^2} ~.
\eea

There has been significant effort invested in developing slow-roll models
of inflation in string theory.  Some fairly recent reviews with further
references are \cite{LindeR,ClineR}.

\subsection {DBI inflation} \label{DBIinflation}
DBI
inflation \cite{Silverstein:2003hf,Alishahiha:2004eh,Chen:2004gc,Chen:2005ad}
is motivated by brane
inflationary
models \cite{Dvali:1998pa,Dvali:2001fw,Burgess:2001fx,Shiu:2001sy,KKLMMT} in
warped
compactifications \cite{Herman,Gukov:1999ya,Dasgupta:1999ss,Greene:2000gh,Giddings:2001yu,Kachru:2003aw}.
In particular, strongly warped regions or
``warped throats'' with exponential warp factors, can arise when there
are fluxes supported on cycles localized in small regions of the
compactified space. A prototypical example of such a strongly warped throat
is the warped deformed conifold \cite{Klebanov:2000hb,Klebanov:2000nc}.
The effective field theory of compact models containing such throats
\cite{Giddings:2001yu,Kachru:2003aw} has been explored in detail in
\cite{WarpedEFT}.
In the slow-roll paradigm, inflation can happen when a brane is
approaching anti-branes in a throat if the potential is flat enough.
However, this is non-generic \cite{KKLMMT}.
Both the degree of tuning involved, and
various
possible ways of engineering flat potentials, have been discussed
in the literature
\cite{KKLMMT,Kallosh,Tye,DeWolfe,Iizuka,TyeII,Cline,Igor}.

Perhaps the most interesting idea, which relies upon dynamics distinct from
the usual slow-roll paradigm, arises in the DBI model.  In this model,
the warped space
slows down the rolling of the
inflaton on even a steep potential.
(This ``slowing down''
can also be understood as arising due to interactions between
the inflaton and the strongly coupled large-N dual field theory).
This scenario can naturally arise in
warped string compactifications \cite{Chen:2004gc}.
The inflaton $\phi$ is the position of a D-brane moving in a warped
throat. In the region where the
back-reaction \cite{Silverstein:2003hf,Chen:2005ad,Chen:2004hu} and
stringy physics \cite{Chen:2005ad,Chen:2005fe} can be ignored, the
effective action has the following form
\begin{equation} \label{Silver}
S=\frac{M_{pl}^2}{2}\int d^4x\sqrt{-g}R - \int d^4x
\sqrt{-g}~ [f(\phi)^{-1}
\sqrt{1+f(\phi)g^{\mu\nu} \partial_\mu \phi \partial_\nu \phi}
-f(\phi)^{-1}+V(\phi)] ~.
\end{equation}
The above expression applies for D3-branes in a warped background where
$f(\phi)$ is the warping factor. We will first express the results in
terms of a general $f(\phi)$. For an AdS-like throat,
$f(\phi)\simeq \frac{\lambda}{\phi^4}$
(where $\lambda$ in specific string constructions
is a parameter which depends 
on the flux numbers).\footnote{This is a good approximation if 
we assume that the last 60 e-foldings of inflation
occur far from the tip of the throat. Otherwise, inflationary
observables may depend on the details
of the warp factor \cite{KMSU}.}
Two situations have been considered in the literature:
\begin{itemize}
\item In the UV model \cite{Silverstein:2003hf,Alishahiha:2004eh}, the
inflaton moves from the UV side of the
warped space to the IR side under the potential
\begin{equation}
V(\phi)\simeq \half m^2 \phi^2 ~, ~~~~ m \gg \Mpl/\sqrt{\lambda} ~.
\end{equation}
In this case the inflaton starts far away from the origin and
rolls relativistically to the minimum of potential at the origin.

\item In the IR model \cite{Chen:2004gc,Chen:2005ad}, the
inflaton moves from the IR side of the
warped space to the UV side under the potential
\begin{equation}
V(\phi)\simeq V_0-\frac{1}{2}m^2\phi^2 ~, ~~~~ m\sim H ~.
\end{equation}
The inflaton starts near the origin and rolls relativistically
away from it.

The evolution of the inflaton in both cases was studied and
the resulting power spectra were
computed in \cite{Silverstein:2003hf,Alishahiha:2004eh,Chen:2004gc,Chen:2005ad}.
Stages of DBI and slow-roll inflation can also be smoothly connected
to each other \cite{Chen:2004gc,Shandera:2006ax}.
\end{itemize}

In the following we summarize the basic results of DBI inflation,
following 
\cite{Silverstein:2003hf,Alishahiha:2004eh,Chen:2004gc,Chen:2005ad},
and using the general formalism developed in \cite{Garriga:1999vw}.
For the zero mode evolution, we assume the
inflaton $\phi$ is spatially homogeneous and denote
$X=\dot{\phi}^2/2$. The pressure $P$ and  energy $E$ are
\begin{eqnarray} \label{energy}
P &=& - f(\phi)^{-1} \sqrt{1- 2X f(\phi)}
+f(\phi)^{-1} - V(\phi)~, \nonumber \\
E &=& 2X P_{,X} -P=
\frac{f(\phi)^{-1}}{\sqrt{1-2X f(\phi)}}-f(\phi)^{-1}+V(\phi)
~,
\end{eqnarray}
and the speed of sound $c_s$
\begin{equation}
c_s=\sqrt{1-\dot{\phi}^2f(\phi)} ~.
\end{equation}

In DBI inflation,
the scalar rolls relativistically and a speed
limit can be inferred by requiring positivity of the argument of
the square root in the
DBI action. So in this limit $c_s\ll 1$, we can approximate the inflaton
speed during inflation by
\begin{equation} \label{speed}
\dot{\phi}\simeq
\pm \frac{1}{\sqrt{f(\phi)}}=\pm \frac{\phi^2}{\sqrt{\lambda}} ~.
\end{equation}

It is easy to see that the requirement $\epsilon \ll 1$ (or
equivalently $|E+P|/E \ll 1$) implies that 
the potential energy $V(\phi)$ dominates throughout inflation
despite the fact that
$\phi$ is rolling relativistically. 
Hence, the Friedmann equation
and the continuity equation reduce to
\begin{eqnarray}
H^2 &=&\frac{V(\phi)}{3 M_{pl}^2} ~,\nonumber \\
V^{\prime}(\phi)&=& -3H\frac{1}{c_s\sqrt{f(\phi)}} ~,
\end{eqnarray}
where we have used the universal speed limit relation
(\ref{speed}) in the continuity equation. The number of e-foldings
is computed as the following
\begin{equation}
N_e=\int_{t_i}^{t_f} H dt=\int_{\phi_i}^{\phi_f}
H\frac{d\phi}{\dot{\phi}}=\int_{\phi_i}^{\phi_f}
\frac{d\phi}{M_{pl}}~\sqrt{\frac{f(\phi)V(\phi)}{3}} ~.
\end{equation}
The scalar power spectrum and gravitational power spectrum are
computed in the general formalism to be
\begin{eqnarray}
P^{\zeta}_k &=&\frac{1}{36 \pi^2
M_{pl}^4}\frac{E^2}{c_s(P+E)}=\frac{f(\phi)V(\phi)^2}{36
\pi^2 M_{pl}^4 } ~, \nonumber \\
P^{h}_k &=& \frac{2E}{3\pi^2 M_{pl}^4}=\frac{2V(\phi)}{3\pi^2
M_{pl}^4} ~.
\end{eqnarray}
The spectral indices
and the running can
also be computed using (\ref{lnkphi}),
\begin{eqnarray} \label{expression}
n_s-1&=&\frac{d \ln P^{\zeta}_k}{d \ln k}
=\frac{\sqrt{3}M_{pl}\phi^2}{\sqrt{\lambda
V}} \left( -\frac{4}{\phi}+\frac{2V^{\prime}}{V} \right) ~,
\nonumber \\
\frac{d n_s}{d\ln
k}&=&\frac{3M_{pl}^2\phi^2}{\lambda} \left(-\frac{4}{V}+\frac{8\phi
V^{\prime}}{V^2}+\frac{2\phi^2V^{\prime\prime}}{V^2}
-\frac{3\phi^2V^{\prime 2}}{V^3} \right) ~,
\nonumber \\
n_T&=&\frac{d \ln P^h_k}{d \ln
k}=\sqrt{\frac{3M_{pl}^2}{\lambda}}\frac{\phi^2V^{\prime}}{V^{\frac{3}{2}}}~,
\end{eqnarray}
where we have evaluated $f(\phi) = \lambda/\phi^4$.
Using the equations of motion it is easy to verify that the
gravitational wave spectral index satisfies the generalized
consistency constraint $\frac{P^h_k}{P^{\zeta}_k}=-8c_s n_{T}$
in \cite{Garriga:1999vw}.

We make two remarks regarding the result (\ref{expression}).
Firstly, as pointed out in \cite{Alishahiha:2004eh}, for the UV model,
both the variation
in the speed of sound and the small parameters $\epsilon$, $\eta$
contribute to the scalar spectral index and their effects cancel
each other in the case of a quadratic potential
$V(\phi)=\frac{1}{2}m^2\phi^2$, so that the spectral index is a
second order quantity $O(\epsilon^2)$ in this case. Indeed, we can
directly see the cancellation from the first formula in
(\ref{expression}) for a quadratic potential. Secondly, 
in the IR model the potential remains
$V(\phi)\simeq V_0$ during inflation, so the above expressions for the
number of e-foldings, the scalar spectral index and its running
(\ref{expression}) become simplified
\begin{eqnarray}
N_e&=&\int_{\phi_i}^{\phi_f}
\frac{d\phi}{\phi^2}~\frac{1}{M_{pl}}\sqrt{\frac{\lambda
V_0}{3}}\simeq \frac{1}{M_{pl}\phi_i}\sqrt{\frac{\lambda
V_0}{3}} ~, \nonumber \\
n_s - 1&=&\frac{\sqrt{3}M_{pl}\phi_i^2}{\sqrt{\lambda
V_0}}(-\frac{4}{\phi_i})=-\frac{4}{N_e} ~,\nonumber \\
\frac{d n_s}{d\ln
k}&=&\frac{3M_{pl}^2\phi_i^2}{\lambda}(-\frac{4}{V_0})=-\frac{4}{N_e^2}
~.
\end{eqnarray}
In this IR model the
gravitational wave production is very much suppressed compared to
the UV model. This suppression is due to the consistency
relation $\frac{P^h_k}{P^{\zeta}_k}=-8c_s n_T$, and to the fact that the
gravitational wave spectral index $n_T$ is much smaller than the scalar
spectral index $n_s-1$ in this case since
$|\frac{V^{\prime}}{V}|\ll |\frac{4}{\phi}|$.

A concern in DBI inflation is how to get the large background charge
$\lambda$ ($\sim 10^{14}$) which is needed to fit the field theory result to 
the observed density perturbations. 
Since this requirement just arises from requiring the compactification
scale in the throat to be $\sim M_{GUT}$ (combined with the standard
AdS/CFT relation between $g_s N$ and the compactification volume), 
it seems very likely that
model building could significantly reduce the apparent tune.
For discussions of this issue, see
\cite{Alishahiha:2004eh,Chen:2005ad,Chen:2005fe}.

\subsection{Kinetically driven inflation}

One simple class of models which can give rise to large non-Gaussianities
is the models of K-inflation, where the dynamics of inflation is governed
by (non-standard) inflaton kinetic terms \cite{Kinflation,Garriga:1999vw}.
The Lagrangians giving rise to K-inflation are not radiatively stable,
and so this mechanism is UV sensitive.
There are as yet no convincing limits of string theory which give rise
to K-inflation, but because the models are so simple, we analyze them in
detail nonetheless.  It would be very interesting to find controlled
limits of string theory which give rise to such models.

The simplest class of K-inflation models are the models of ``power-law
K-inflation."
The Lagrangian for power-law K-inflation is of the form
\begin{equation}
\label{klag}
P(X,\phi) = {4\over 9}{{(4-3\gamma)}\over \gamma^2}
{1\over \phi^2}(-X + X^2) ~,
\end{equation}
where $\gamma$ is a constant, not to be confused with the Lorentz factor 
($1/c_s$) that often appears in the literature of DBI inflation.
(The form of the Lagrangian and our discussion 
can be straightforwardly generalized to a more
arbitrary form where $P \propto f(X)/\phi^2$ \cite{Kinflation}.)
Before describing the physics which follows from (\ref{klag}), we should
discuss some general concerns about K-inflation.
The most obvious (also mentioned above)
is that a Lagrangian of the form (\ref{klag}) is not
radiatively stable, since it is not protected by any symmetry.  (A shift
symmetry of $\phi$ could protect a Lagrangian of the form $P(X)$ with
generic coefficients).
A second concern is
that a reasonable exit mechanism for inflation must be provided.
A third, related concern is that the dominant energy
condition
\begin{equation}
\label{dec}
{\partial P \over \partial X} \geq 0,~~X {\partial P \over \partial X} -
P \geq 0
\end{equation}
is not satisfied by (\ref{klag}) for small values of $X$.  Therefore,
while in the inflating solution we will see that (\ref{dec}) is satisfied,
one must provide an exit mechanism that changes the form of $P$ drastically
enough that the physics around flat space is sensible.
We shall discuss these issues further after summarizing the key properties
of the
solution of interest.

One solution to the equations of motion \cite{Kinflation}
is to take
\begin{equation}
\label{xis}
X ~=~X_0~= {{2-\gamma} \over {4 - 3\gamma}}
\end{equation}
which gives rise to an FRW cosmology with
\begin{equation}
a(t) \sim t^{{2\over 3\gamma}}
\end{equation}
for any $0 < \gamma < 2/3$.
The speed of sound following from (\ref{klag}) is
\begin{equation}
\label{cis}
c_s^2 = {{\gamma}\over {8-3\gamma}}~.
\end{equation}
We are most interested in the regime with $c_s << 1$.
Therefore, we will focus on models with small $\gamma$, and sometimes
expand formulae around $\gamma \to 0$.

\subsubsection{The effective theory governing small fluctuations}

To get some intuition for these models, it is useful to construct an
effective theory describing small fluctuations around the background
inflating solution.  The equations (\ref{xis}) imply that
\begin{equation}
\phi(t) \sim (1 + \gamma/8) t
\end{equation}
where we have absorbed an overall constant into the definition of $t$, and
have only written the solution to ${\cal O}(\gamma)$.  Let us cast the
Lagrangian into a more familiar form by performing the field redefinition
\begin{equation}
\Phi = {\rm log}(\phi)
\end{equation}
valid for $t>0$.  Then the mini-superspace Lagrangian takes the form
\begin{equation}
\label{newlis}
{\cal L} ~=~f(\gamma)\left(-{1\over 2}\dot \Phi^2 +
{1\over 4}\dot \Phi^4 e^{2\Phi} \right)
\end{equation}
where $f(\gamma)$ is the complicated $\gamma$-dependent prefactor in
(\ref{klag}); for small $\gamma$
\begin{equation}
\label{fis}
f(\gamma) \sim {16\over 9\gamma^2}~.
\end{equation}
Defining $Y = -{1\over 2} g^{\mu\nu}\partial_{\mu}\Phi \partial_{\nu}\Phi$,
the full Lagrangian is just
\begin{equation}
\label{ylag}
{\cal L} = f(\gamma) \left(-Y + Y^2 e^{2\Phi}\right)~.
\end{equation}

Now, we introduce the effective field $\pi$ which describes small fluctuations
around the solution via
\begin{equation}
\Phi(x,t) = \pi(x,t) + \Phi_0(t)
\end{equation}
where $\Phi_0$ characterizes the inflationary solution, with
\begin{equation}
\label{solnis}
\Phi_0(t) = {\rm log} \left((1+{\gamma \over 8}) t\right)
\end{equation}
(and therefore
$Y_0 = {1\over 2}{1\over t^2}$).  We will also find it useful to
define
\begin{equation}
Z = -{1\over 2}g^{\mu\nu}\partial_{\mu}\pi \partial_{\nu}\pi~.
\end{equation}

Expanding ${\cal L}$ around this solution, we find a Lagrangian for
$\pi$ of the form
\begin{equation}
\label{piLis}
{\cal L}(\pi,\dot\pi,\nabla \pi) = f(\gamma) \left(\tilde {\cal L}_0 + \tilde
{\cal L}_2 + \tilde {\cal L}_3 \right)~.
\end{equation}
Here, the subscript on the ${\tilde {\cal L}}$ denotes its order in the
fluctuation $\pi$.  The first order term is guaranteed to vanish in the
background $\Phi_0$, since it solves the equations of motion.
To compute the Lagrangian up to third order in $\pi$, we need the expansions
of $Y$ and $e^{2\Phi}$.
Using the explicit solution (\ref{solnis})
we see that
\begin{equation}
Y = {1\over 2t^2} + {1\over t} \dot \pi + Z
\end{equation}
and
\begin{equation}
e^{2\Phi} = t^2 (1 + {\gamma \over 4}) \left( 1 + 2\pi + 2 \pi^2 +
{4\over 3}\pi^3 + \cdots \right)~.
\end{equation}
We then find
\begin{equation}
\label{Lzero}
\tilde{\cal L}_0 = -{1\over 4}{1\over t^2} (1-{\gamma\over 4})~,
\end{equation}
\begin{equation}
\label{Ltwo}
\tilde{\cal L}_2 = (1 + {\gamma\over 4}) \dot \pi^2 +
{\gamma \over 4}Z + {\pi^2 \over 2t^2} (1 + {\gamma \over 4})
+ 2 {\pi \dot \pi \over t}(1 + {\gamma\over 4})~,
\end{equation}
and
\begin{equation}
\label{Lthree}
\tilde{\cal L}_3 = (1 + {\gamma\over 4}) \left( {1\over 3}{\pi^3 \over t^2}
+ 2{\pi^2 \dot \pi \over t} + 2\pi Z + 2\pi \dot\pi^2 + 2t \dot \pi Z
\right) ~.
\end{equation}
Here, the field $\pi$ has dimension one, and all
terms should be rendered dimension four by appropriate powers of $\Mpl$.
One can use the relation $H \sim {2\over 3\gamma t}$ to replace explicit
powers of $t$ above with powers of $H$ and $\gamma$, in estimating the
size of various terms.

\subsubsection{Basic phenomenology}

Here, we describe the rough phenomenology of a `realistic' model of power
law K-inflation.  The exit from inflaton will be discussed in a later
subsection.

The power spectrum of these models was derived by Garriga and Mukhanov
\cite{Garriga:1999vw}.  Working in the limit of small $\gamma$, their answer
(equation (42) of \cite{Garriga:1999vw}) becomes
\begin{equation}
\label{scapower}
P^{\zeta}_k = {1\over c_s}{2\over 3\gamma} {G H_1^2 \over \pi} \left(
{k\over k_1}\right)^{-3\gamma}
= {1\over c_s}{2\over 3\gamma} {H_1^2 \over 8\pi^2 M_{pl}^2} \left( {k\over k_1}
\right)^{-3\gamma}
\end{equation}
where $H_1$ is taken to be the Hubble scale at the time of horizon exit for
the perturbations currently at our horizon, and $k_1$ is the the associated
comoving wavenumber.

It follows from (\ref{scapower}) that the tilt
\begin{equation}
n_s - 1~=~-3 \gamma + \cdots
\end{equation}
which allows us to fix $\gamma \sim 1/60$ using the central
value of the spectral index in the WMAP results
\cite{WMAP}.
This justifies our use of perturbation theory in $\gamma$ in earlier
equations.

Using (\ref{cis}) and (\ref{scapower}), as well as the fact that
data determines $P^{\zeta} \sim 10^{-9}$ at horizon crossing, we find
\begin{equation}
\label{solvH}
H^2 \sim {3\over 4\sqrt{2}} \gamma^{3/2} 8 \pi^2 M_{pl}^2 \times 10^{-9}
\end{equation}
at horizon crossing, where $M_{pl}$ is the reduced Planck mass.
Plugging in $\gamma \sim {1\over 60}$ as determined from $n_s$, we see that the Hubble
scale when the 60th from the last e-folding leaves our horizon is
roughly $H \sim 10^{-5} M_{pl}$.
So primordial gravitational waves will be unobservable in this model.

The reader may wonder about the following.  In these models, with
$c_s << 1$, the sound horizon where fluctuations freeze out at $c_s/H$ can
be much smaller than $1/H$.
What happens if ${c_s \over H} < l_p$, i.e. $H > c_s M_{pl}$?  This would seem
to give rise to a ``trans-Planckian problem."

However, it is easy to see that this regime cannot be reached in any reliable
fashion.  The power spectrum (\ref{scapower}) makes it clear that
\begin{equation}
{\delta \rho \over \rho} \sim {1\over \gamma^{3/4}}\left(H \over M_{pl} \right)~.
\end{equation}
Hence, for $H > \gamma^{1/2} M_{pl}$, ${\delta \rho \over \rho} \geq {\cal O}(1)$
and there is no good semi-classical description of any resulting region
which exits from inflation. 

\subsubsection{Exiting from K-inflation}

Using $a(t) \sim t^{2\over 3\gamma}$, the
requirement that one gets 60 e-foldings starting from some initial time
$t_i$ is simply
\begin{equation}
{\rm log}(t_f/t_i) ~=~90 \gamma ~.
\end{equation}
Then, we need to arrange for an appropriate exit mechanism to kick in
at the $t_f$ we determine in this way.

Here, we briefly describe a simple mechanism to exit from K-inflation,
modeled on hybrid inflation \cite{Hybrid}.
Going back to the Lagrangian (\ref{ylag}), we would like to arrange
so that after 60 e-foldings, at some specific value of $\phi$, the
inflationary stage ends and the exit to standard radiation domination
occurs.  An easy way to do this, while fixing the problem that (\ref{ylag})
violates the DEC around $X=0$, is to consider a more elaborate theory
including also a second scalar field $\psi$.  Then a Lagrangian of the
schematic form
\begin{equation}
\label{exitlag}
{\cal L} = -Y + Y^2 e^{2\Phi} + (\partial \psi)^2 + (\Phi_*^2 - \Phi^2) \psi^2
+ {1\over M_*^2} \psi^2 Y + \alpha (\psi^2 - \beta^2 M_*^2)^2 + \cdots
\end{equation}
can do the trick.
Here $M_*$ is some UV scale, perhaps the string scale or
the Planck scale.
For small $\alpha  \beta^2$,
assuming $\Phi_* \sim M_*$, then
right around the time when $\langle \Phi \rangle \sim \Phi_*$
the $\psi$ field becomes
tachyonic and condenses.  It rolls to a vev $\langle \psi \rangle
\sim \beta M_*$, and for $\beta \geq {\cal O}(1)$, can correct the
sign of the $\phi$ kinetic term.

The Lagrangian (\ref{exitlag}) can then support an early phase of
K-inflation, and exit to a phase with normal kinetic terms for the
various fields.  The DEC is satisfied during both the inflationary
phase and around the flat-space vacuum with $\langle \psi \rangle =
\beta M_*$.  It is an interesting question to check whether it is
satisfied all along the trajectory from the inflationary phase to
the final vacuum.

\section{Non-Gaussian Perturbations}
\setcounter{equation}{0}

Now in the general setup of Sec.~\ref{SecSetup}, we consider the
non-Gaussian perturbations in the primordial power spectrum. There
is a large literature on this subject, see
e.g. \cite{Maldacena:2002vr,Seery:2005wm,Gangui:1993tt,Acquaviva:2002ud,Alishahiha:2004eh,Gruzinov:2004jx,Chen:2005fe,Calcagni:2004bb}.
However, most of the literature has
been focused on the case where the speed of sound $c_s$ is very close to
one, where the primordial non-Gaussianities are generally too small
to be detected in future experiments.  In addition, in some of
the literature, only the perturbations in the
matter Lagrangian are considered, but not the gauge invariant
perturbation that remains exactly constant after horizon exit.\footnote{For
models with significant non-Gaussianity, this may be a reasonable
approximation, since the contributions from the gravitational sector
yield an $f_{NL}$
which is too small to measure in any case.}
Here we will consider a
general Lagrangian of the form (\ref{general}), and we will allow
the speed of sound $c_s$ to assume arbitrary values, only requiring the
parameters in (\ref{small}) (and one more parameter to be defined
later) to be small and of order
$\CO(\epsilon)$. We will calculate the three-point correlation
function for the gauge invariant scalar perturbation $\zeta$ following
the approach of Maldacena \cite{Maldacena:2002vr}.

It is useful to define two parameters following \cite{Seery:2005wm}
\begin{eqnarray}
\Sigma&=&X P_{,X}+2X^2P_{,XX}  = \frac{H^2\epsilon}{c_s^2} ~,\\
\lambda&=& X^2P_{,XX}+\frac{2}{3}X^3P_{,XXX} ~.
\label{lambda}
\end{eqnarray}
In the case of inflationary models with $P_{,X\phi}=0$, the
parameter $\lambda$ can be written in terms of the speed of sound
and small parameters $\epsilon$, $\eta$, $s$ in (\ref{small}).
However, for inflation models where $P_{,X\phi}\neq 0$, such as
DBI inflation or K-inflation, there is no simple formula for the
parameter $\lambda$ in terms of the slow variation parameters, and we must
treat each model individually.

To compute the Einstein action to the third
order, it is useful to work in the ADM metric formalism
\begin{equation}
ds^2=-N^2dt^2+h_{ij}(dx^i+N^i dt)(dx^j+N^j dt) ~.
\end{equation}
This formalism is convenient because the equations of motion for
the variables $N$ and $N^i$ are quite easy to solve. We will work
in a comoving gauge where the three dimensional metric $h_{ij}$
takes the form
\begin{equation}
h_{ij}=a^2 e^{2\zeta}\delta_{ij} ~,
\end{equation}
where we have neglected the tensor perturbations. $a$ is the scale
factor of the universe and $\zeta$ is
the scalar perturbation, and remains constant outside the horizon in
this gauge.
The index on $N^i$ can be lowered by the 3-dim metric $h_{ij}$.
The inflaton
fluctuation $\delta \phi$ vanishes in this gauge, which makes the
computations simpler. Using the ADM metric ansatz the action
becomes
\begin{equation} \label{action}
S=\frac{1}{2}\int dt d^3x \sqrt{h} N (R^{(3)}+2P)+\frac{1}{2}\int
dt d^3x \sqrt{h} N^{-1}(E_{ij}E^{ij}-E^2)
\end{equation}
where we have set the reduced Planck mass $M_{pl}=1$ for
convenience. The three-dimensional Ricci curvature $R^{(3)}$ is
computed from the metric $h_{ij}$. The symmetric tensor $E_{ij}$ is
defined as
\begin{equation}
E_{ij}=\frac{1}{2}(\dot{h}_{ij}-\nabla_i N_j-\nabla_j N_i) ~.
\label{Edef}
\end{equation}
The equations of motion for $N$ and $N^i$ are
\begin{eqnarray} \label{eomN}
R^{(3)}+2P-4X P_{,X}-N^{-2}(E_{ij}E^{ij}-E^2)=0 ~, \nonumber \\
\nabla_j(N^{-1}E_i^j)-\nabla_i(N^{-1}E)=0 ~.
\end{eqnarray}

We follow \cite{Maldacena:2002vr} and decompose $N^i$ into two parts
$N_i=\tilde{N}_i+\partial_i\psi$ where $\partial_i\tilde{N}^i=0$, 
and expand $N$ and $N^i$ in powers of $\zeta$
\begin{eqnarray}
N&=&1+\alpha_1+\alpha_2+\cdots ~, \nonumber \\
\tilde{N}_i&=&N^{(1)}_i+N^{(2)}_i+\cdots ~, \nonumber \\
\psi&=&\psi_1+\psi_2+\cdots ~,
\end{eqnarray}
where $\alpha_n, \tilde{N}^{(n)}_i, \psi_n \sim O(\zeta^n)$. One can
plug the power expansion into the equations of motion (\ref{eomN})
for $N$ and $N^i$. At first order in $\zeta$, the
solutions \cite{Maldacena:2002vr,Seery:2005wm} are
\begin{eqnarray} \label{solution1}
\alpha_1 = \frac{\dot \zeta}{H}~,~~~
N^{(1)}_i=0 ~,~~~
\psi_1=-\frac{\zeta}{H}+\chi ~, ~~~
\partial^2 \chi= a^2 \frac{\epsilon}{c_s^2} \dot \zeta ~,
\end{eqnarray}
after choosing proper boundary conditions.

In order to compute the effective action to order $\CO(\zeta^3)$,
as pointed out in \cite{Maldacena:2002vr}, in the ADM
formalism one only needs to consider the
perturbations of $N$ and $N^i$ to the first order $\CO(\zeta)$. This
is because their perturbations at order
$\CO(\zeta^{3})$ such as $\alpha_3$ will multiply the
constraint equation
at the zeroth order $\CO(\zeta^0)$ which vanishes, and the second
order perturbations such as $\alpha_2$ will
multiply a factor which vanishes by the first
order solution (\ref{solution1}). So the solution
(\ref{solution1}) is enough for our purpose. This general conclusion
in the ADM formalism can be seen as follows.

What we have done so far is solve the constraint equations for the
Lagrange multipliers $N$ and $N_i$, which result from the variation of
the action with respect to them
\bea
\delta S &=& \int d^4x ~\delta \CL (\partial_i N, N) \cr
&=& \int d^4x \left[ \frac{\partial \CL}{\partial(\partial_i N)}
\partial_i \delta N + \frac{\partial \CL}{\partial N} \delta N \right]
~,
\label{variation}
\eea
where for simplicity we schematically denote $N$ as either $N$ or
$N_i$.
We expand $N = N^{(0)} + \Delta N = N^{(0)} + N^{(1)} + N^{(2)} +
\cdots$ for
\bea
\frac{\partial \CL}{\partial(\partial_i N)}
&=& \frac{\partial \CL}{\partial(\partial_i N)} |_0 +
\frac{\partial^2 \CL} {\partial(\partial_i N)\partial(\partial_j N)}|_0
\partial_j \Delta N + \cdots ~, \nonumber \\
\frac{\partial \CL}{\partial N}
&=& \frac{\partial \CL}{\partial N}|_0 +
\frac{\partial^2 \CL}{\partial N^2}|_0 \Delta N + \cdots ~,
\eea
where the subscript 0 means $\Delta N=0$. In order to get $N^{(1)}$ we can
neglect the terms
involving $\frac{\partial^2 \CL}{\partial(\partial_i N) \partial
N}$. This is
because this term starts from $\CO(\zeta)$ as we can see from
(\ref{action}) and (\ref{Edef}), so it does not contribute to $\Delta
N$ at the first order $\CO(\zeta)$.

The $\CO(\zeta^0)$ terms in (\ref{variation}) are
\bea
\int d^4x \left[ \frac{\partial
\CL}{\partial(\partial_i N)}|_{0,\zeta^0} \partial_i \delta N
+ \frac{\partial \CL}{\partial N}|_{0,\zeta^0} \delta N \right] =0 ~.
\label{constraint0}
\eea
This equation is consistent with the background equation of motion
(\ref{eomH}). The $\CO(\zeta)$ terms in (\ref{variation}) are
\bea
\int d^4x \left\{ \left[ \frac{\partial \CL}{\partial(\partial_i N)}
|_{0,\zeta}
+ \frac{\partial^2 \CL}{\partial(\partial_i N) \partial(\partial_j N)}
|_{0,\zeta^0} \partial_j N^{(1)} \right] \partial_i \delta N
\right. \cr
\left. + \left[ \frac{\partial \CL}{\partial N}|_{0,\zeta}
+ \frac{\partial^2 \CL}
{\partial N^2}|_{0,\zeta^0} N^{(1)} \right] \delta N \right\} =0 ~,
\label{constraint1}
\eea
where the subscripts $\zeta$ or $\zeta^0$ denote the order of the
perturbation $\zeta$
that we take.
After integration by parts this gives the constraint equations for
$\Delta N$ at order $\CO(\zeta)$, namely $N^{(1)}$.
A similar procedure can be used
to solve for $\Delta N$ to order $\CO(\zeta^n)$, namely up to $N^{(n)}$.

We will next substitute these solutions for the Lagrange multipliers
into the action and expand to order $\CO(\zeta^n)$, where $n\geq 3$. 
We show that to do
this the knowledge up to $N^{(n-2)}$ is enough. Let us look at the
terms that possibly contain $N^{(n-1)}$ and $N^{(n)}$,
\bea
\Delta S = \int d^4x \left\{ \frac{\partial
\CL}{\partial(\partial_i N)}|_0 \partial_i \Delta N
+ \half
\frac{\partial^2 \CL}{\partial(\partial_i N)\partial(\partial_j N)}|_0
(\partial_i \Delta N)(\partial_j \Delta N)
\right. \cr \left.
+ \frac{\partial \CL}{\partial N}|_0 \Delta N
+ \half \frac{\partial^2 \CL}{(\partial N)^2}|_0 (\Delta N)^2 + \cdots
\right\} ~.
\label{Sexpansion}
\eea
The terms involving $\frac{\partial^2 \CL}{\partial(\partial_i
N)\partial N} = \CO(\zeta) $ are not written because such terms will
not contain $N^{(n-1)}$ or $N^{(n)}$ at order $\CO(\zeta^n)$.

The following are all terms containing $N^{(n)}$ $(n\geq 2)$ in
(\ref{Sexpansion}),
\bea
\int d^4x \left\{ \frac{\partial
\CL}{\partial(\partial_i N)}|_{0,\zeta^0} \partial_i N^{(n)}
+ \frac{\partial \CL}{\partial N}|_{0,\zeta^0} N^{(n)} \right\} ~.
\eea
Comparing with (\ref{constraint0}) we know that
this term vanishes. This is because after integration by parts, $N^{(n)}$ will
multiply a term which is just the zeroth order constraint equation
coming from
(\ref{constraint0}) after integration by parts.
Next we look at all terms containing $N^{(n-1)}$ $(n\geq 3)$  in
(\ref{Sexpansion}),
\bea
\int d^4x \left\{ \frac{\partial \CL} {\partial(\partial_i N)}
|_{0,\zeta} \partial_i N^{(n-1)} + \frac{\partial^2 \CL}
{\partial(\partial_i N)\partial(\partial_j N)} |_{0,\zeta^0}
\partial_i N^{(n-1)} \partial_j N^{(1)}
\right. \cr \left.
+ \frac{\partial \CL}{\partial N}|_{0,\zeta} N^{(n-1)} +
\frac{\partial^2 \CL}{\partial N^2}|_{0,\zeta^0} 
N^{(n-1)} N^{(1)} \right\} ~.
\eea
This term also vanishes because, after integration by parts,
$N^{(n-1)}$
will multiply a term which is the first order constraint equation
coming from
(\ref{constraint1}) after integration by parts.

Therefore our task is simplified. In order to expand the action
(\ref{action}) to 
quadratic and cubic order in the  
primordial scalar perturbation $\zeta$, we only need to plug in the
solution for the first
order perturbation in $N$ and $N^i$ and do the expansion.
The results can also be extracted\footnote{Note that there is a typo
in
$\frac{1}{a^4}\frac{\dot{\cal R}}{H}$ term in Eq (44) of
\cite{Seery:2005wm}, it
should be $\frac{1}{a^4}\frac{\dot{\cal R}}{H} \partial^2 \psi_1
\partial^2 \psi_1$  instead.}
from \cite{Seery:2005wm,Maldacena:2002vr}
\begin{eqnarray}
S_2 &=& \int dt d^3x~
[a^3  \frac{\epsilon}{c_s^2}\dot\zeta^2- a \epsilon
(\partial \zeta)^2 ] ~,
\label{actionQuad} \\
S_3&=&\int dt d^3x [ -\epsilon a \zeta(\partial \zeta)^2
-a^3 (\Sigma+2\lambda) \frac{\dot{\zeta}^3}{H^3} +
\frac{3a^3\epsilon}{c_s^2} \zeta \dot \zeta^2
\nonumber \\
&+& \frac{1}{2a} (3\zeta-\frac{\dot{\zeta}}{H})
(\partial_i\partial_j\psi\partial_i\partial_j\psi-\partial^2\psi\partial^2\psi)
-2 a^{-1} \partial_i\psi\partial_i\zeta\partial^2\psi ] ~,
\label{actionCubic}
\end{eqnarray}
where $\dot{\zeta}$ is the derivative with respect to $t$. One can
decompose the perturbations into momentum modes using
\begin{eqnarray}
u_k=\int d^3x~
\zeta(t,\textbf{x})e^{-i\textbf{k}\cdot \textbf{x}} ~.
\end{eqnarray}

\subsection{The quadratic part}
\label{SecQuad}
To solve the quadratic part of the action (\ref{actionQuad}) we
define
\bea
v_k\equiv z u_k ~, ~~~~z \equiv \frac{a \sqrt{2\epsilon}}{c_s} ~.
\label{vdef}
\eea
This brings the equation of motion for the perturbation $\zeta$ to a
simple form
\bea
v_k'' + c_s^2 k^2 v_k - \frac{z''}{z} v_k =0 ~,
\label{quadeom}
\eea
where the prime denotes the derivative with respect to the conformal
time defined by $dt=ad\tau$, $\tau = -(aH)^{-1} (1 + \CO(\epsilon))$.
The leading order of $z''/z = 2a^2 H^2 (1+\CO(\epsilon))$ is
contributed by
the scale factor $a$ which has the strongest time dependence.
If the sound speed varies slowly enough, the leading behavior of the
equation (\ref{quadeom}) is given by a Bessel function.
We write it in terms of the Fourier modes of $\zeta$, $u_k$, using
(\ref{vdef}),
\begin{eqnarray}
u_k= u(\tau,{\bf k}) = \frac{i H}{\sqrt{4\epsilon
c_s k^3}}(1+i k c_s\tau)e^{-i k c_s\tau} ~.
\label{uk}
\end{eqnarray}
Here we have made the approximation that the sound speed changes
slowly, so this solution has oscillatory behavior but
with a frequency that is slowly changing due to the time dependence of
$c_s$. This requires
\bea
-k \tau \Delta c_s \ll c_s k \Delta \tau ~.
\eea
That is, the phase change in (\ref{uk}) caused by the change of
$c_s$
is much slower than that caused by the change of $\tau$. This
condition can be brought to the form
\bea
\frac{\dot c_s}{c_s} \ll H ~,
\eea
which is just the condition for small slow variation parameter $s$
in (\ref{small}).

From Eq.~(\ref{uk}) we can see that,
before horizon exit $c_s k > aH$, $u_k$ is oscillating and its
amplitude is decreasing proportionally to $\tau$. For $v_k$, this is
the leading
behavior in flat space and we have chosen the standard Bunch-Davies
vacuum. (We will discuss non-Gaussianities for other choices of vacua
in Section \ref{Nonstandard}).
After the horizon
exit $c_s k < aH$, $u_k$ remains constant \cite{Mukhanov:1990me}. This
is most easily seen
from (\ref{quadeom}) where we can neglect the second term, and we see
that (for the growing mode)
$v_k \propto z_k$ so that $u_k = {\rm constant}$. This conclusion is
not going to be changed by the higher order
interactions \cite{Maldacena:2002vr}, because, as
we can see from the interaction terms (\ref{actionCubic}), they involve
either spatial derivatives, which can be neglected at super-horizon
scales, or powers of time derivatives starting from second order.
So after horizon
exit the leading value of $u_k$ is determined by (\ref{uk}) at $\tau
\approx 0$ with the rest of the variables evaluated at $c_s k = a H$.
We emphasize here that the validity of our analysis
only requires the ${\it variation}$ of sound
speed to be slow; the sound speed can be arbitrary. For our later
purposes, the first order corrections to the leading behavior
(\ref{uk}) is also important. We work this out in Appendix
\ref{SecukCorr}.

Now, we follow the standard technique in quantum field theory and write
the operator in terms of creation and annihilation modes
\begin{eqnarray}
\zeta(\tau,\textbf{k})=u(\tau,\textbf{k})a(\textbf{k})
+u^*(\tau,-\textbf{k})a^{\dagger}(-\textbf{k})
\end{eqnarray}
with the canonical commutation relation
$[a(\textbf{k}),a^{\dagger}(\textbf{k}^{\prime})]
=(2\pi)^3\delta^{(3)}(\textbf{k}-\textbf{k}^{\prime})$.

\subsection{The cubic part}
\label{SecCubic}
The cubic effective action in (\ref{actionCubic}) looks like order
$\CO(\epsilon^0)$ in the slow variation parameters. In slow-roll
inflation, as
emphasized and demonstrated in Ref.~\cite{Maldacena:2002vr},
one can perform a
lot of integrations by parts and cancel terms of order
$\CO(\epsilon^0)$ and $\CO(\epsilon)$. The resulting cubic action is
actually of leading order $\CO(\epsilon^2)$ in slow roll parameters.
A similar analysis can be performed for the general
Lagrangian in Ref.~\cite{Seery:2005wm}, as well as
in the case of interest here where the sound speed is arbitrary. 
Except for terms that
are proportional to $1-c_s^2$ or $\lambda$, the rest of the terms can
be cancelled to the second order $\CO(\epsilon^2)$,
\begin{eqnarray} \label{action3}
S_3&=&\int dt d^3x\{
-a^3 (\Sigma(1-\frac{1}{c_s^2})+2\lambda)\frac{\dot{\zeta}^3}{H^3}
+\frac{a^3\epsilon}{c_s^4}(\epsilon-3+3c_s^2)\zeta\dot{\zeta}^2
\nonumber \\ &+&
\frac{a\epsilon}{c_s^2}(\epsilon-2s+1-c_s^2)\zeta(\partial\zeta)^2-
2a \frac{\epsilon}{c_s^2}\dot{\zeta}(\partial
\zeta)(\partial \chi) \nonumber \\ &+&
\frac{a^3\epsilon}{2c_s^2}\frac{d}{dt}(\frac{\eta}{c_s^2})\zeta^2\dot{\zeta}
+\frac{\epsilon}{2a}(\partial\zeta)(\partial
\chi) \partial^2 \chi +\frac{\epsilon}{4a}(\partial^2\zeta)(\partial
\chi)^2+ 2 f(\zeta)\frac{\delta L}{\delta \zeta}|_1 \} ~,
\end{eqnarray}
where the variable $\chi$ is defined in Eq.~(\ref{solution1}),
and in the last term
\begin{eqnarray}
\frac{\delta
L}{\delta\zeta}\mid_1 &=& a
\left( \frac{d\partial^2\chi}{dt}+H\partial^2\chi
-\epsilon\partial^2\zeta \right) ~,
\end{eqnarray}
\begin{eqnarray} \label{redefinition}
f(\zeta)&=&\frac{\eta}{4c_s^2}\zeta^2+\frac{1}{c_s^2H}\zeta\dot{\zeta}+
\frac{1}{4a^2H^2}[-(\partial\zeta)(\partial\zeta)+\partial^{-2}(\partial_i\partial_j(\partial_i\zeta\partial_j\zeta))] \nonumber \\
&+&
\frac{1}{2a^2H}[(\partial\zeta)(\partial\chi)-\partial^{-2}(\partial_i\partial_j(\partial_i\zeta\partial_j\chi))] ~.
\end{eqnarray}
Here $\partial^{-2}$ is the inverse Laplacian,
$\frac{\delta
L}{\delta\zeta}|_1$ is the variation of the
quadratic action with respect to the perturbation $\zeta$, therefore
the last term which is proportional to $\frac{\delta
L}{\delta\zeta}|_1$ can be absorbed by a field redefinition of
$\zeta$. It can be easily shown that the field redefinition that
absorbs this term is
\begin{eqnarray}
\zeta \rightarrow \zeta_n+f(\zeta_n) ~.
\end{eqnarray}
For the correlation function only the first term in
(\ref{redefinition})
contributes since all other terms involve at least one derivative
of $\zeta$ that vanish outside the horizon. The
three-point function after field redefinition $\zeta \rightarrow
\zeta_n+\frac{\eta}{4c_s^2}\zeta_n^2$ becomes
\begin{eqnarray}
\langle\zeta(\textbf{x}_1)\zeta(\textbf{x}_2)\zeta(\textbf{x}_3)\rangle
&=&\langle\zeta_n(\textbf{x}_1)\zeta_n(\textbf{x}_2)\zeta_n(\textbf{x}_3)\rangle
\nonumber \\
&+&\frac{\eta}{2c_s^2}
(\langle\zeta_n(\textbf{x}_1)\zeta_n(\textbf{x}_2)\rangle
\langle\zeta_n(\textbf{x}_1)\zeta_n(\textbf{x}_3)\rangle+\textrm{sym})
+ \CO (\eta^2 (P^\zeta_k)^3) ~.
\end{eqnarray}

We proceed to calculate with the above cubic terms
(\ref{action3}). The terms in the last line in (\ref{action3}) are all
of subleading order in slow variation parameters (\ref{small}). The
interaction Hamiltonian from the leading terms to $\CO(\epsilon^2)$ is
\begin{eqnarray} \label{action4}
H_{int}(t)&=&-\int
d^3x\{ -a^3 (\Sigma(1-\frac{1}{c_s^2})+2\lambda)\frac{\dot{\zeta}^3}{H^3}+
\frac{a^3\epsilon}{c_s^4}(\epsilon-3+3c_s^2)\zeta\dot{\zeta}^2
\nonumber \\ &+&
\frac{a\epsilon}{c_s^2}(\epsilon-2s+1-c_s^2)\zeta(\partial\zeta)^2-2a\frac{\epsilon}{c_s^2}\dot{\zeta}(\partial
\zeta)(\partial \chi) \}~.
\end{eqnarray}
One then computes the vacuum expectation value of the three point
function in the interaction picture that characterizes the
primordial non-Gaussianities
\begin{eqnarray}
\langle
\zeta(t,\textbf{k}_1)\zeta(t,\textbf{k}_2)\zeta(t,\textbf{k}_3)\rangle=
-i\int_{t_0}^{t}dt^{\prime}\langle[
\zeta(t,\textbf{k}_1)\zeta(t,\textbf{k}_2)\zeta(t,\textbf{k}_3),H_{int}(t^{\prime})]\rangle ~,
\end{eqnarray}
where $t_0$ is some very early time when the vacuum fluctuation of
the inflaton is well within the horizon, and $t$ is a time about
several e-foldings after the horizon exit. Translated to the
conformal time $\tau=-1/(a H)$, we can in a good
approximation take the integral over conformal time $\tau$ from
$-\infty$ to $0$. We follow the standard procedure and compute the
contributions from various terms. In the following we first evaluate
the leading contributions of each term.

\begin{enumerate}
\item Contribution from $\dot{\zeta}^3$ term. We denote
$K=k_1+k_2+k_3$, and find
\begin{eqnarray}
&& -i(c_s^2-1+\frac{2\lambda
c_s^2}{\Sigma})\frac{H^2\epsilon}{c_s^4}u(0,\textbf{k}_1)
u(0,\textbf{k}_2)u(0,\textbf{k}_3)
\int
_{-\infty}^{0}\frac{a d\tau}{H^3} \nonumber \\ &\times&
(6\frac{d u^*(\tau,\textbf{k}_1)}{d \tau}\frac{d
u^*(\tau,\textbf{k}_2)}{d \tau}\frac{d u^*(\tau,\textbf{k}_3)}{d
\tau})
(2\pi)^3 \delta^3(\sum_i\bk_i) + {\rm c.c.}
\nonumber \\
&=& -\frac{3H^4}{8\epsilon^2c_s^4}(c_s^2-1+\frac{2\lambda
c_s^2}{\Sigma})(2\pi)^3\delta^{3}(\textbf{k}_1+\textbf{k}_2+\textbf{k}_3)
(\prod_{i=1}^{3}\frac{1}{k_i^3}) (\frac{k_1^2k_2^2k_3^2}{K^3}) ~.
\label{IntFirst}
\end{eqnarray}

\item Contribution from $\zeta\dot{\zeta}^2$ term. We find
\begin{eqnarray}
&&i\frac{\epsilon}{c_s^4}(\epsilon-3+3c_s^2)u(0,\textbf{k}_1)
u(0,\textbf{k}_2)u(0,\textbf{k}_3) \int
_{-\infty}^{0} a^2 d\tau \nonumber \\ &\times&
2(u^*(\tau,\textbf{k}_1)\frac{d u^*(\tau,\textbf{k}_2)}{d
\tau}\frac{d u^*(\tau,\textbf{k}_3)}{d \tau}+\textrm{sym})
(2\pi)^3 \delta^3(\sum_i\bk_i) +{\rm c.c.}
\nonumber \\
&=& \frac{H^4}{16\epsilon^2c_s^4}(\epsilon -3+
3c_s^2)(2\pi)^3\delta^{3}(\textbf{k}_1+\textbf{k}_2+\textbf{k}_3)
(\prod_{i=1}^{3}\frac{1}{k_i^3}) \nonumber \\ &\times&
(\frac{k_2^2k_3^2}{K}+\frac{k_1k_2^2k_3^2}{K^2}+\textrm{sym}) ~.
\label{IntSecond}
\end{eqnarray}

\item Contribution from $\zeta(\partial \zeta)^2$ term.
\begin{eqnarray}
&& \frac{H^4}{16\epsilon^2c_s^4}(\epsilon -2s+1-
c_s^2)(2\pi)^3\delta^{3}(\textbf{k}_1+\textbf{k}_2+\textbf{k}_3)
 (\prod_{i=1}^{3}\frac{1}{k_i^3})\nonumber \\ &\times&
((\textbf{k}_1\cdot \textbf{k}_2)
(-K+\frac{k_1k_2+k_1k_3+k_2k_3}{K}+\frac{k_1k_2k_3}{K^2})+\textrm{sym})
~. 
\label{IntThird}
\end{eqnarray}

\item Contribution from $\dot{\zeta}(\partial
\zeta)(\partial \chi)$ term.
\begin{eqnarray}
&& -\frac{H^4}{16\epsilon
c_s^4}(2\pi)^3\delta^{3}(\textbf{k}_1+\textbf{k}_2+\textbf{k}_3)
 (\prod_{i=1}^{3}\frac{1}{k_i^3})\nonumber \\ &\times&
(\frac{(\textbf{k}_1\cdot
\textbf{k}_2)k_3^2}{K}(2+\frac{k_1+k_2}{K}) +\textrm{sym}) ~.
\end{eqnarray}

\item Contribution from field redefinition $\zeta \rightarrow
\zeta_n+\frac{\eta}{4c_s^2}\zeta_n^2$.
\begin{eqnarray}
&& \frac{\eta}{2}\frac{H^4}{16\epsilon^2
c_s^4}(2\pi)^3\delta^{3}(\textbf{k}_1+\textbf{k}_2+\textbf{k}_3)
 (\frac{1}{k_1^3k_2^3}
+\textrm{sym}) ~.
\end{eqnarray}

\end{enumerate}

As a first step, we add all these leading contributions together.
After some simplification,
we find
\begin{eqnarray}
\langle \zeta(\textbf{k}_1)\zeta(\textbf{k}_2)\zeta(\textbf{k}_3)\rangle
&=&
(2\pi)^7\delta^3(\textbf{k}_1+\textbf{k}_2+\textbf{k}_3)
(P_k^\zeta)^2
\frac{1}{\prod_i k_i^3}~ \CA ~,
\label{3point}
\end{eqnarray}
where the above contributions to $\CA$ are organized as follows
\begin{eqnarray}
\CA &\supset& \left(\frac{1}{c_s^2}-1
-\frac{2\lambda}{\Sigma} \right) \frac{3k_1^2k_2^2k_3^2}{2K^3} 
\nonumber \\
&+&
\left(\frac{1}{c_s^2}-1\right)
\left(-\frac{1}{K}\sum_{i>j}k_i^2k_j^2+\frac{1}{2K^2}
\sum_{i\neq j}k_i^2k_j^3+\frac{1}{8}\sum_{i}k_i^3 \right) 
\nonumber \\
&+& \frac{\epsilon}{c_s^2} \left( -\frac{1}{8}
\sum_i k_i^3 + \frac{1}{8}
\sum_{i\neq j} k_i k_j^2 + \frac{1}{K} \sum_{i>j} k_i^2 k_j^2
\right) 
\nonumber \\
&+& \frac{\eta}{c_s^2} \left( \frac{1}{8} \sum_i k_i^3
\right)
\nonumber \\
&+& \frac{s}{c_s^2} \left( -\frac{1}{4} \sum_i k_i^3 - \frac{1}{K}
\sum_{i>j} k_i^2 k_j^2 +
\frac{1}{2K^2} \sum_{i\neq j} k_i^2 k_j^3 \right) ~.
\label{3pointPre}
\end{eqnarray}

\subsection{Correction terms}
\label{SecCorr}
We notice that, in (\ref{3pointPre}) for general $c_s$, the first two
lines are not of the same order of magnitude as the last three
lines. The former are $\CO(1)$ while the latter are $\CO(\epsilon)$. 
So for $c_s^2 \ll 1$, 
it is clear that the first two terms dominate.  If for completeness
one is interested 
in the full result to $\CO(\epsilon)$, however, 
small corrections to the first two lines may compete with the last three
lines.  
This means that one
must 
calculate the
subleading terms (of order $\CO(\epsilon)$) 
for the first three integrations in
Sec.~\ref{SecCubic}.

In obtaining (\ref{3pointPre}), we treat all the slow-varying
parameters in the integrand as constant, and we use the leading order
solution (\ref{uk}). 
The corrections come from several sources. 

Firstly, there are
corrections to the leading order $u(\tau, k)$ in (\ref{uk}), which we
work out in Appendix \ref{SecukCorr},
\bea
u_k(y) = -\frac{\sqrt{\pi}}{2\sqrt{2}} ~\frac{H}{\sqrt{\epsilon c_s}}~
\frac{1}{k^{3/2}} (1+\frac{\epsilon}{2} +\frac{s}{2} )
~e^{i \frac{\pi}{2} (\epsilon + \frac{\eta}{2}) }~ y^{3/2}
H_{\frac{3}{2}+\epsilon+\frac{\eta}{2}+\frac{s}{2}}^{(1)} 
\left( (1+\epsilon+s)y \right) ~,
\eea
where $y = \frac{c_s k}{a H}$.

Secondly, various parameter in this solution as well as others in the
integrand, including $H$, $c_s$, $\lambda$, $\epsilon$, are all
time-dependent subject to the slow variation conditions. We Taylor
expand such functions as 
\bea
f(\tau) &=& f(t_K) + \frac{\partial f}{\partial t} (t-t_K) +
\CO (\epsilon^2 f) \nonumber \\
&=& f(\tau_K) - \frac{\partial f}{\partial t} \frac{1}{H_K} \ln
\frac{\tau}{\tau_K} + \CO(\epsilon^2 f)
~,
\eea
where the reference point $\tau_K$ is chosen to be the moment when the
wave-number $K=k_1+k_2+k_3$ exits the horizon.

Thirdly, the scale factor $a$ also receives $\CO(\epsilon)$
correction,
\bea
a = -\frac{1}{H_K \tau} - \frac{\epsilon}{H_K \tau} 
+ \frac{\epsilon}{H_K \tau} 
\ln (\tau/\tau_K) + \CO(\epsilon^2) ~.
\eea

We then consider all types of corrections in the first three
integrations (\ref{IntFirst}), (\ref{IntSecond}) and (\ref{IntThird}) 
in Sec.~\ref{SecCubic}.
We leave the details of these calculations to Appendix \ref{AppCorr},
and summarize the final results in the following subsection.

\subsection{Summary of final results}
\label{SecSumFinal}
To  first order in the slow variation parameters $\CO(\epsilon)$,
the three-point correlation function of the gauge invariant scalar
perturbation $\zeta$ for a general single field inflation model is
given by the following:
\begin{eqnarray}
\langle \zeta(\textbf{k}_1)\zeta(\textbf{k}_2)\zeta(\textbf{k}_3)\rangle
&=&
(2\pi)^7\delta^3(\textbf{k}_1+\textbf{k}_2+\textbf{k}_3)
(\tilde P_K^\zeta)^2
\frac{1}{\prod_i k_i^3} \cr &\times&
(\CA_\lambda +{\CA}_c + \CA_o +\CA_\epsilon +\CA_\eta +\CA_s) ~,
\label{3pointFinal}
\end{eqnarray}
where we have decomposed the shape of the three point function
into six parts
\begin{eqnarray}
\CA_\lambda &=& \left(\frac{1}{c_s^2}-1 
- \frac{\lambda}{\Sigma}[2- (3-2\bc_1)l] \right)_K
\frac{3k_1^2k_2^2k_3^2}{2K^3} ~, 
\label{Alam} \\
\CA_c &=&
\left(\frac{1}{c_s^2}-1\right)_K
\left(-\frac{1}{K}\sum_{i>j}k_i^2k_j^2+\frac{1}{2K^2}
\sum_{i\neq j}k_i^2k_j^3+\frac{1}{8}\sum_{i}k_i^3 \right) ~, 
\label{Ac} \\
\CA_o &=& 
\left(\frac{1}{c_s^2}-1 - \frac{2\lambda}{\Sigma} \right)_K
\left(\epsilon F_{\lambda\epsilon} + \eta F_{\lambda\eta} + s F_{\lambda s}
 \right)
\nonumber \\
&+& \left(\frac{1}{c_s^2}-1 \right)_K
\left(\epsilon F_{c\epsilon} + \eta F_{c\eta} + s F_{cs} \right) ~,
\label{Ao} \\
\CA_\epsilon &=& \epsilon \left( -\frac{1}{8}
\sum_i k_i^3 + \frac{1}{8}
\sum_{i\neq j} k_i k_j^2 + \frac{1}{K} \sum_{i>j} k_i^2 k_j^2
\right) ~, 
\label{Aep} \\ 
\CA_\eta &=& \eta \left( \frac{1}{8} \sum_i k_i^3
\right)~,
\label{Aeta} \\
\CA_s &=& s F_s ~.
\label{As}
\end{eqnarray}
The definitions of the sound speed $c_s$, $\Sigma$ and $\lambda$ are
\bea
c_s^2 &\equiv& \frac{P_{,X}}{P_{,X} + 2X P_{,XX}} ~,
\nonumber \\
\Sigma &\equiv& X P_{,X} + 2X^2 P_{,XX} ~,
\nonumber \\
\lambda &\equiv& X^2 P_{,XX} + \frac{2}{3} X^3 P_{,XXX} ~.
\eea
The definitions of the four slow variation parameters are
\bea
\epsilon \equiv -\frac{\dot H}{H^2} ~, ~~~~
\eta \equiv \frac{\dot \epsilon}{\epsilon H} ~, ~~~~
s \equiv \frac{\dot c_s}{c_s H} ~, ~~~~
l \equiv \frac{\dot \lambda}{\lambda H} ~.
\label{slowvariation}
\eea
$\tilde P^\zeta_K$ is defined as 
\bea
\tilde P^\zeta_K \equiv \frac{1}{8\pi^2} \frac{H_K^2}{c_{s K}\epsilon_K} ~.
\label{tP}
\eea
Note that $H$, $c_s$, $\epsilon$, 
$\lambda$ and $\Sigma$ in this final result are 
evaluated at the moment $\tau_K \equiv -\frac{1}{Kc_{s K}} +\CO(\epsilon)$
when the wave number $K\equiv k_1 +k_2 +k_3$ exits the horizon
$K c_{s K} = a_K H_K$, as indicated by the subscript $K$. 
Various $F$'s are functions of $k_i$, whose
detailed
forms are given in Appendix \ref{AppCorrFinal}.

These results clearly illustrate that very significant non-Gaussianities
$f_{NL} \gg 1$ will arise most easily in models with $c_s \ll 1$ or
$\frac{\lambda}{\Sigma} \gg 1$ during
inflation (while conventional slow-roll models enjoy $c_s = 1$ and
$\lambda/\Sigma =0$).
We also see that for this wide
class of models, the functional form of the
leading non-Gaussianity is completely determined in terms
of 5 numbers: $c_s$, ${\lambda\over\Sigma}$, and the three slow variation
parameters $\epsilon$, $\eta$ and $s$. Note that in
(\ref{3pointPre})
we are assuming that these parameters do not vary over
the few e-foldings we see close to the horizon.  If they do,
the parameter counting becomes a little more complicated, and 
one can in
principle extract more information from these results (by studying
the running of the non-Gaussianities). Indeed after Taylor-expanding
some slow-varying functions in working out all the correction
terms in Sec.~\ref{SecCorr}, one more parameter $l$ comes up. However
as illustrated in (\ref{Alam}), it happens that one can absorb $l$ in 
$\frac{2\lambda}{\Sigma}$ as $\frac{\lambda}{\Sigma}
[2-(3-2\bc_1)l]$, where ${\bf c}_1$ is the Euler constant. 
(Later we will often refer to it as
$\frac{2\lambda}{\Sigma}$ for simplicity.)
The error introduced to (\ref{3pointFinal}) after this absorption is
of order $\CO(\epsilon^2)$.
So to the first order $\CO(\epsilon)$ that we
are interested, our final result is still parameterized by five
numbers.

\section{Size, shape and running of the non-Gaussianities}
\label{SecSSR}
\setcounter{equation}{0}

In the previous section we have obtained the most general form of the
primordial three-point scalar non-Gaussianities up to first order in slow
variation parameters (\ref{slowvariation}) in single field
inflationary models,
where the matter Lagrangian is an arbitrary function of the inflaton
and its first derivative. This non-Gaussianity is controlled by five
parameters --- three
small parameters $\epsilon$, $\eta$ and $s$, the sound speed
$c_s$ and another parameter $\lambda/\Sigma$. There is a large
literature studying non-Gaussian features in models belonging
to this class. One interesting feature of our result is that we can
take different limits and smoothly connect various previous
results. We will also explore regions which have not been studied
before. Before giving
several major examples, we first discuss some general features of the
non-Gaussianity obtained in Sec.~\ref{SecSumFinal}.

The correlation function in Sec.~\ref{SecSumFinal} is a
function of three momenta forming a triangle. Therefore generally
there are three interesting properties --- the magnitude of the
function, its dependence on
the shape of the triangle and its dependence on the size of the
triangle. Namely, these quantities determine the size, shape and running of the
non-Gaussianity.

To discuss whether a non-Gaussianity is large enough to be observed we
first need to quote the experimental sensitivities. The non-Gaussianity
of the CMB in the WMAP observations is analyzed by assuming the
following ansatz for the scalar perturbation
\bea
\zeta = \zeta_L - \frac{3}{5} f_{NL} \zeta_L^2 ~,
\eea
where $\zeta_L$ is the linear Gaussian part the perturbations, and
$f_{NL}$ is an estimator parameterizing the size of the
non-Gaussianity.\footnote{The sign convention of $f_{NL}$ here follows
  Ref.~\cite{Maldacena:2002vr}, and it is opposite to the WMAP's
  sign convention.}
This assumption leads to the following three-point correlation function
\begin{eqnarray}
\langle \zeta(\textbf{k}_1)\zeta(\textbf{k}_2)\zeta(\textbf{k}_3)\rangle
=(2\pi)^7\delta^3(\textbf{k}_1+\textbf{k}_2+\textbf{k}_3)(-\frac{3}{10}f_{NL}(P^{\zeta}_k)^2)\frac{\sum_i k_i^3}{\prod_i k_i^3} ~.
\end{eqnarray}
Notice that this shape is different from any of the shapes
in Sec.~\ref{SecSumFinal} except for $\CA_\eta$. But we
can set up a similar
estimator $f_{NL}$ for each of those different shapes of
non-Gaussianities to parameterize its
magnitude. This matching is conventionally done for the equilateral
triangle case $k_1=k_2=k_3$. We then have
\bea
f_{NL}^\lambda &=& -\frac{5}{81}
\left(\frac{1}{c_s^2}-1-\frac{2\lambda}{\Sigma} \right) 
+ (3-2\bc_1) \frac{l \lambda}{\Sigma} ~,
\nonumber \\
f_{NL}^c &=& \frac{35}{108} \left(\frac{1}{c_s^2}-1 \right) ~,
\nonumber \\
f_{NL}^o &=& \CO \left( \frac{\epsilon}{c_s^2} ~,~
\frac{\epsilon\lambda}{\Sigma} \right) ~,
\nonumber \\
f_{NL}^{\epsilon,\eta,s} &=& \CO (\epsilon) ~.
\label{fNL5}
\eea
The bound on the parameter $f_{NL}$ from data analysis depends on the
shape of the non-Gaussianities which we will discuss shortly. 
The current bound is roughly $|f_{NL}|
< 300$ for the first three shapes \cite{Creminelli:2005hu}, and
$|f_{NL}| < 100$ for
the rest \cite{WMAP}. A non-Gaussianity is potentially
observable in future experiments if $|f_{NL}| >5$
\cite{Komatsu:2001rj,Komatsu:2003fd,Verde:1999ij}.

The magnitudes of $\CA_\epsilon$, $\CA_\eta$ and $\CA_s$ are unobservably
small, of order $\CO(\epsilon)$.
The sizes of $\CA_o$ are determined
by $\epsilon/c_s^2$, $\eta/c_s^2$ and $s/c_s^2$. 
So in order to make the magnitude of
these functions larger, we need the denominator $c_s^2$ to be less
than at
least one of the slow variation parameters. 
(A similar conclusion for $\lambda/\Sigma$ term.)
It is very
interesting to construct such models, and we will show an example in
this section. The most significant non-Gaussianities come from
$\CA_\lambda$ and $\CA_c$ when $c_s \ll 1$ 
and/or $\lambda/\Sigma \gg 1$.

\begin{figure}
\begin{center}
\epsfig{file=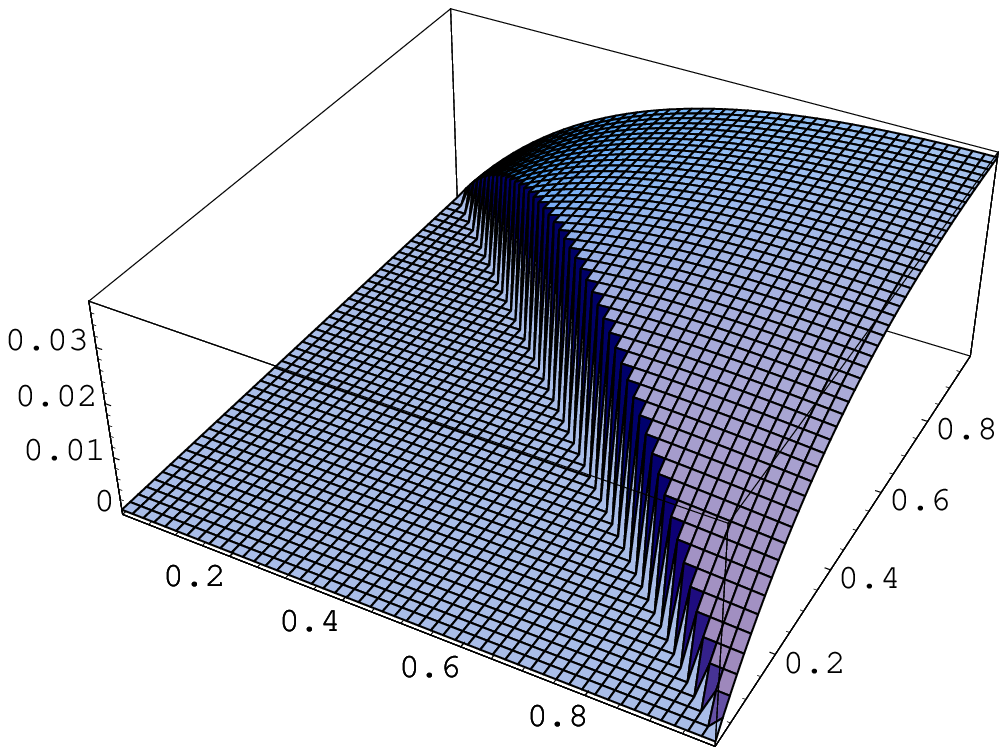, width=10cm}
\end{center}
\medskip
\caption{The shape of $-\CA_\lambda/k_1k_2k_3$}
\label{Flam}
\end{figure}

\begin{figure}
\begin{center}
\epsfig{file=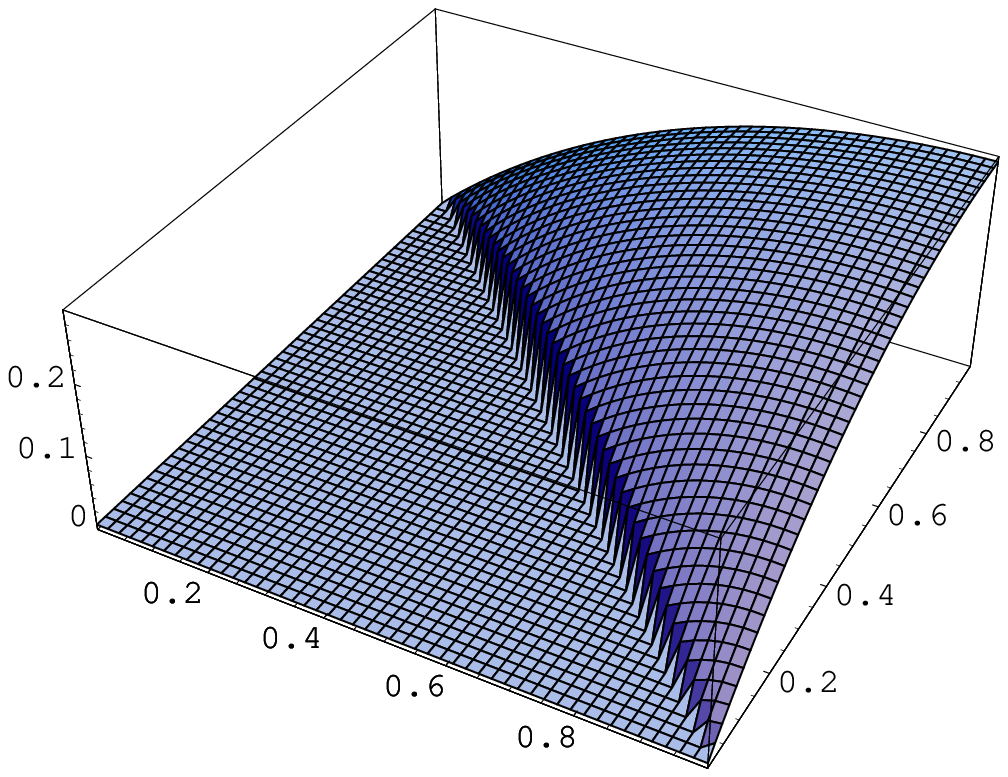, width=10cm}
\end{center}
\medskip
\caption{The shape of $\CA_c/k_1k_2k_3$}
\label{Fc}
\end{figure}

\begin{figure}
\begin{center}
\epsfig{file=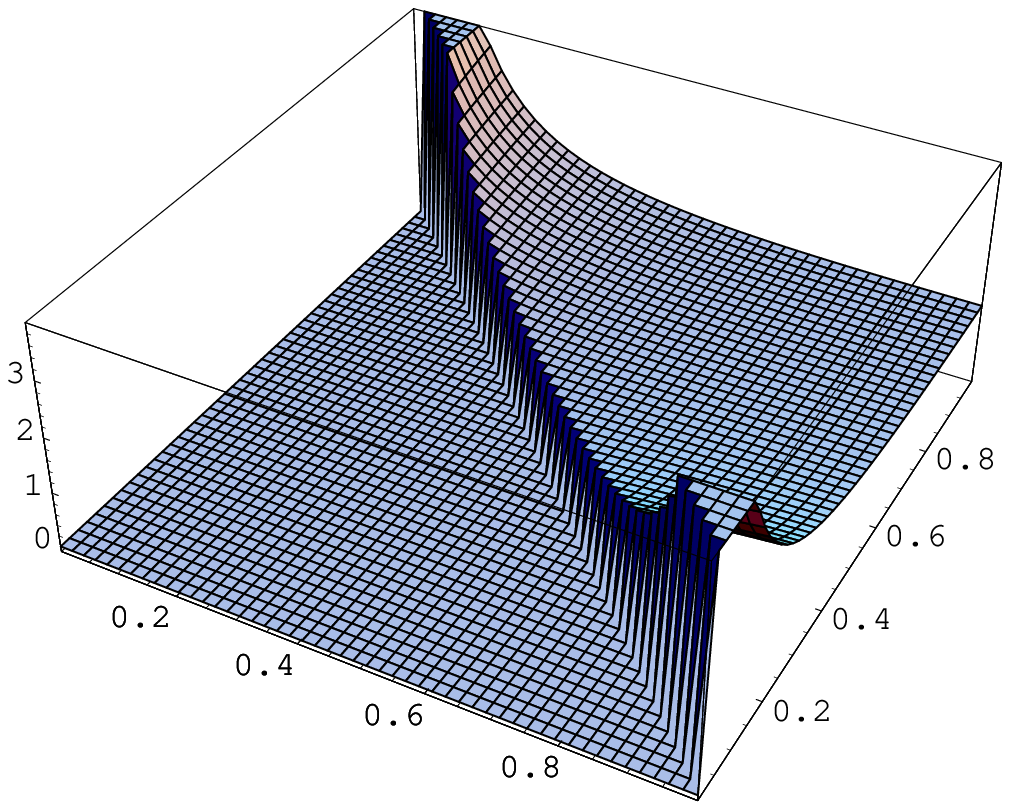, width=10cm}
\end{center}
\medskip
\caption{The shape of $-\CA_\epsilon/k_1k_2k_3$}
\label{Fep}
\end{figure}

As in
Ref.~\cite{Babich:2004gb}, to show the shape of $\CA(k_1,k_2,k_3)$, we
present the 3-d plot $x_2^{-1} x_3^{-1}
\CA(1,x_2,x_3)$ as a function of $x_2=k_2/k_1$ and $x_3=k_3/k_1$. 
The shapes of $-\CA_\lambda/k_1k_2k_3$ and 
$\CA_c/k_1k_2k_3$ are shown in Fig.~\ref{Flam} and \ref{Fc}.
We can see that $\CA_\lambda$ and $\CA_c$ have overall
similar shapes, but with opposite sign.
The shape of $-\CA_\epsilon/k_1k_2k_3$ is shown in
Fig.~\ref{Fep}. The shapes of
$\CA_\eta$, $\CA_s$ are similar to that of $\CA_\epsilon$
up to a sign, in the sense that they all roughly approach to 
a pole in the squeezed limit, e.g. when $k_1=k_2$ and $k_3 \to
0$. 

For a small sound speed, both the leading
non-Gaussianity $\CA_\lambda$, $\CA_c$ and the subleading
non-Gaussianity $\CA_o$ are potentially observable. 
Their shapes in the squeezed limit are all similar,
\bea
\frac{\CA_{\lambda,c,o}}{k_1 k_2 k_3} \propto \frac{k_3}{k_1} =
 x_3 ~.
\eea
Details are shown in Appendix \ref{SecSqueezed}.

The size of the non-Gaussianity also depends on the scale that we
measure. Analogous to the spectral index, we define \cite{Chen:2005fe}
\bea
n_{NG}-1 \equiv \frac{d \ln |f_{NL}|}{d \ln k} ~.
\eea
For example, if the main contribution to $f_{NL}$ comes from a small
sound speed, then $n_{NG}-1 \approx -2s$. So in this case a measurement of
the running of the non-Gaussianity directly tells us one of the slow
variation parameter $s$.

\subsection{Slow-roll inflation}

We now reduce Eq.~(\ref{3pointFinal}) to the slow-roll case. In this case
the deviation of the sound speed from one is very small. We denote
$u=1-\frac{1}{c_s^2} \ll 1$. Assuming $u=\CO(\epsilon)$, we can
neglect both $\CA_s$ (\ref{As}) since $s \approx \frac{\dot u}{2H}
\ll \CO(\epsilon)$ and $\CA_o$ since $\CA_o = \CO(u\epsilon)$. 
As in Ref.~\cite{Seery:2005wm}, if we further
assume $P_{,X\phi} = 0$, the relations
\bea
\lambda = \frac{\epsilon}{6} \left(\frac{2\epsilon}{3\epsilon_X}(1-u)s
- u \right) ~, ~~~~ \epsilon_X \equiv -\frac{\dot X}{H^2}
\frac{\partial H}{\partial X} 
\eea
follow.
In this limit the Eq.~(\ref{Alam})-(\ref{As}) become
\bea
\CA &=& -\left(\frac{\epsilon}{3\epsilon_X}s+u \right)
\frac{k_1^2k_2^2k_3^2}{K^3}  \nonumber \\
&-& u
\left(-\frac{1}{K}\sum_{i>j}k_i^2k_j^2+\frac{1}{2K^2}
\sum_{i\neq j}k_i^2k_j^3+\frac{1}{8}\sum_{i}k_i^3 \right) 
\nonumber \\ &+&
\epsilon \left( -\frac{1}{8}
\sum_i k_i^3 + \frac{1}{8}
\sum_{i\neq j} k_i k_j^2 + \frac{1}{K} \sum_{i>j} k_i^2 k_j^2
\right) 
\nonumber \\ &+&
\eta \left( \frac{1}{8} \sum_i k_i^3 \right)~.
\eea
Taking into account of the convention difference $\CA_{\rm SL} =
4\CA_{\rm ours}$, we recover the result Eq.(83) of Seery and Lidsey.

Further setting the sound speed to be 1 (therefore $u=0$ and $s=0$),
using the relation (\ref{smallrelation}) and taking into account of
the convention difference $\CA_{\rm Maldacena} = 8\CA_{\rm ours}$, we
recover the result Eq.(4.6) of Maldacena.

\subsection{DBI inflation}
\label{SecDBI}
In Ref.~\cite{Gruzinov:2004jx,Alishahiha:2004eh,Chen:2005fe}, the
three-point correlation functions in various models with
potentially larger non-Gaussianities have been calculated by
considering the perturbations in the matter Lagrangian only.  
In general such a procedure is not valid because it is only the
gauge invariant combination $\zeta$ that will remain constant after
the horizon exit. However (as realized by these authors) 
for the large non-Gaussianity case, due to
large non-linear self coupling of the inflaton, the
corrections coming from the gravitational part are relatively small,
and this simpler method can be successfully used to obtain the
leading behavior of the non-Gaussianities. Indeed, in this limit our
leading behavior in $\CA_\lambda$
and $\CA_c$, which is of order
$\CO(c_s^{-2},\lambda/\Sigma)$, recovers the leading behavior of the
result
in Eq.(6) of Gruzinov\cite{Gruzinov:2004jx}.
However this procedure does not guarantee that the subleading term in
Ref.~\cite{Gruzinov:2004jx},
which is of $\CO(1)$, will be correct. In fact, as we
discussed at the beginning of this section, for
$0<c_s^2<\epsilon,\eta,s$, the subleading terms actually come from
$\CA_o$ and a subleading term in $\CA_\lambda$, which are of order
$c_s^{-2}\CO(\epsilon,\eta,s)$ and should be observable.

Now we consider the example of DBI inflation
discussed in Sec.~\ref{DBIinflation}. Interestingly for this type of
inflation model,
because of the Lagrangian
(\ref{energy}), the parameter $\lambda$ defined in (\ref{lambda}) is
\begin{equation} \label{DBIlambda}
\lambda=
X^2P_{,XX}+\frac{2}{3}X^3P_{,XXX}=\frac{H^2\epsilon}{2c_s^4}(1-c_s^2)
~.
\end{equation}
So the leading order
contribution in $\CA_\lambda$ vanishes. The leading behavior of $\CA_c$
reproduces the result of Alishahiha, Silverstein and
Tong\cite{Alishahiha:2004eh,Chen:2005fe}.
Using the estimator defined in (\ref{fNL5}), we have
\begin{equation}
f^c_{NL}\approx 0.32\frac{1-c_s^2}{c_s^2} \approx 0.32 c_s^{-2} ~.
\end{equation}
To estimate the subleading order corrections, let us look
at both the UV and IR model. 
In the UV model, the sound speed is given by
\bea
c_s^{-1} \approx \sqrt{\frac{2\lambda}{3}}\frac{m M_{pl}}{\phi^2} ~.
\eea
It is easy to see that whether the subleading order
is of order $\CO(1)$ or $\CO(\epsilon/c_s^2)$ depends on the value of
$\phi$.
In the IR model, the sound speed is related to
the number of  e-foldings $N_e$ before the end of inflation by
\bea
c_s^{-1} \approx \beta N_e/3 ~,
\eea
where $\beta$ parameterizes the steepness of the potential $\beta
\equiv m^2/H^2$, which is generically of order one as we know from the
usual eta-problem in slow-roll inflation. Since $\epsilon, \eta, s =
\CO(N_e^{-1})$, we see that this model falls into the
region where $c_s^2 \ll \CO(\epsilon,\eta,s)$, where both 
the leading non-Gaussianity $\CA_c$ and the subleading $\CA_o$ are
observable.

We can also
compute the running of non-Gaussianities as considered in
\cite{Chen:2005fe}
\begin{eqnarray}
n_{NG}-1
\approx -2s = -2\sqrt{\frac{3
M_{pl}^2}{f(\phi)
V(\phi)}}
\left( \frac{1}{2}\frac{V^{\prime}}{V}
-\frac{V^{\prime\prime}}{V^{\prime}}+\frac{2}{\phi} \right) ~.
\end{eqnarray}
Just like the spectral index, the running non-Gaussianities also
provide a good probe of the effective potential $V(\phi)$ and
warp factor $f(\phi)$, and could (in an optimistic scenario)
distinguish between different
possible background geometries where the brane motion is occurring.

\subsection{Kinetic inflation}

We saw above that in one large class of models with measurable
non-Gaussianities, the leading order of $\CA_{\lambda}$ vanishes.
However, generally it does not vanish for
some other models such as K-inflation, and could in principle be
comparable to the second term $\CA_c$. An experimental
detection of the shape of large non-Gaussianities could
therefore in principle distinguish
between DBI inflation and more general models where the first term
has significant contributions.

Here, we first outline a quick estimate of the non-Gaussianities in
K-inflation, and then describe the exact result following \S4.

\subsubsection{Crude estimate of non-Gaussianities}

Given the fluctuation Lagrangians
(\ref{Ltwo}) and (\ref{Lthree}), we can do a simple estimate of the
non-Gaussianity (following the general strategy also employed in
\cite{ghost,Alishahiha:2004eh}).  While
this is not strictly necessary in view of
the detailed formulae in \S4, it is perhaps illuminating to understand
in simple terms why these models have large $f_{NL}$.

The basic point is the following.  We saw in (\ref{scapower}) that
\begin{equation}
P_{k}^{\zeta} \sim {1\over \gamma^{3/2}} \left( {H \over M_{pl}} \right)^2~.
\end{equation}
Using the fact that
\begin{equation}
\zeta \sim {H\over \dot \Phi} \pi
\end{equation}
and evaluating this on the inflationary solution, we find
\begin{equation}
\zeta \sim {\pi \over \gamma M_{pl}}~.
\end{equation}
It follows that
\begin{equation}
\langle \zeta \zeta \rangle \sim {1\over \gamma^2}\langle \pi \pi \rangle~.
\end{equation}
Furthermore, given the overall factor of $f(\gamma) \sim {1\over \gamma^2}$
in ${\cal L}$, which in particular multiplies the $\pi$ kinetic
terms, one has the relation
\begin{equation}
\langle \pi \pi \rangle \sim \gamma^2 (\delta \pi)^2
\end{equation}
where $\delta \pi$ is a typical fluctuation of the $\pi$ field during
inflation.  Therefore, estimating $P^{\zeta}$ via
\begin{equation}
P^{\zeta} \sim {1 \over \gamma^2} \langle \pi \pi \rangle \sim
(\delta \pi)^2
\end{equation}
and using $P^{\zeta} \sim {H^2\over \gamma^{3/2}}$, we see that
\begin{equation}
\label{pifluct}
\delta \pi \sim {H \over \gamma^{3/4}}~.
\end{equation}
Using the form of the modes one can also see that
\begin{equation}
\label{dotfluct}
\delta \dot \pi \sim H \delta \pi~.
\end{equation}

Now, we are interested in estimating the non-Gaussianity, say through a
crude estimate of $f_{NL}$.  Since naively
\begin{equation}
{\tilde{\cal L}_3 \over {\tilde
{\cal L}_2}}
\sim f_{NL} \sqrt{P^{\zeta}}~,
\end{equation}
we can plug
(\ref{pifluct}) and (\ref{dotfluct})
into the fluctuation Lagrangians to estimate the
non-Gaussianity.
The term $2t\dot \pi Z$ in $\tilde{\cal L}_3$ contributes fluctuations
of size ${1\over M_{pl}} {1\over H\gamma} ({H^2\over \gamma^{3/4}})^3 \sim
{1\over \gamma^{13/4}}(H^5/M_{pl})$.
The largest terms in $\tilde{\cal L}_2$
scale like ${H^4\over \gamma^{3/2}}$.
We then find
\begin{equation}
{\tilde {\cal L}_3 \over \tilde {\cal L}_2} \sim {1\over \gamma^{7/4}}
{H\over M_{pl}}~.
\end{equation}
Because
\begin{equation}
\sqrt{P^{\zeta}} \sim {1\over \gamma^{3/4}} {H\over M_{pl}}
\end{equation}
this translates into the rough estimate
\begin{equation}
\label{guesstimate}
f_{NL} \sim {1\over \gamma} \sim {1\over c_s^2}~.
\end{equation}
This in fact reproduces the more detailed results of \S4 when applied
to K-inflation, though it does
not give the (potentially very important) information about the detailed shape
of the momentum dependence.

\subsubsection{Shape of Non-Gaussianities in K-inflation}
An order of magnitude estimate for the general size of the non-Gaussian
signatures arising in K-inflation, appears in (\ref{guesstimate}).
Here, we refine this estimate using the formulae of \S4.

The Lagrangian of power law K-inflation, at leading order in the
expansion in $\gamma$, is given by
\begin{equation}
P(X,\phi) = {16\over 9\gamma^2} {1\over \phi^2}(-X + X^2)~.
\end{equation}
We then see that
\begin{equation}
\Sigma = {16\over 9\gamma^2}{1\over \phi^2} \left(6X^2 - X\right)
\sim {16\over 9 \gamma^2} {1\over \phi^2}
\end{equation}
and
\begin{equation}
\lambda = {32 \over 9\gamma^2}{X^2 \over \phi^2}
\sim {8 \over 9\gamma^2}{1\over \phi^2}~.
\end{equation}
In each case, the estimate after $\sim$ follows from the fact that
$X = {1\over 2} + {\cal O}(\gamma)$ on the inflationary solution.

We find that, to leading order in $\gamma$,
\begin{equation}
\label{lambdaco}
{\cal A}_{\lambda} = {12\over \gamma} \left({k_1^2 k_2^2 k_3^2\over K^3}
\right)
\end{equation}
and
\begin{equation}
{\cal A}_{c} = {8\over \gamma} \left( -{1\over K} \sum_{i>j} k_i^2 k_j^2
+ {1\over 2K^2} \sum_{i \neq j} k_i^2 k_j^3 + {1\over 8}  \sum_i k_i^3
\right)~.
\end{equation}
So unlike the other higher derivative model we've carefully examined,
DBI inflation, these models receive a leading-order contribution from
the shape ${\cal A}_{\lambda}$. Also, in distinction to a general 
DBI inflation model in Sec.~\ref{SecDBI},
the constant speed of sound implies that the
non-Gaussianity does not run in power law K-inflation.

Evaluating on equilateral triangles and comparing to the ``local'' form of
non-Gaussianities, this translates to an estimated $f_{NL}$ of
\begin{equation}
f_{NL} \approx {170 \over 81}{1\over \gamma}~.
\end{equation}
This gives $f_{NL} \approx 125$ for realistic models, which is allowed by
current experimental bounds but would be ${\it easily}$ detectable in
future experiments.  Comparing to the rough estimate (\ref{guesstimate}),
which was $f_{NL} \sim {8\over \gamma}$, we see that the two results
agree up to an ${\cal O}(1)$ coefficient (whose moderate smallness
prevents the model from being excluded by current data).

\section{Non-Gaussianities as a probe of the inflationary vacuum}
\label{Nonstandard}
\setcounter{equation}{0}

There has been some interest in the question of whether we can
observe trans-Planckian physics in the Cosmic Microwave Background
radiation \cite{Brandenberger,CGS,Shiu,Danielsson,Kempf,Kaloper,Burgess,BEFT,Porrati,CH,Mottola}.
In this context, the assumption that the Bunch-Davies vacuum
is the unique initial state of inflation has recently been
questioned.
While several plausible alternatives for the initial state have been suggested, its precise
form is highly dependent on how we model the behavior of
quantum fields at Planckian energies.
What is universal however is that any
deviation from the
Bunch-Davies vacuum during inflation will
result in modulations of the power spectrum \cite{Shiu,Danielsson,BEFT},
thus offering the exciting possibility of probing the initial state of the universe from
cosmological measurements.

We hasten to stress that the microphysics that determines
the choice of inflationary vacuum is by no means understood.
Here we put aside the conceptual issues associated with the choice of vacuum
and its consistency, and simply approach the problem of vacuum
ambiguity from a phenomenological
perspective. To be specific, we explore the possibility of using
primordial non-Gaussianities to test any deviation from the
standard Bunch-Davies vacuum. As it turns out, the effect of deviation
from the Bunch-Davies vacuum on the shape of the primordial
non-Gaussianities is quite simple to compute within our formalism. A
general vacuum state for the fluctuation of the inflaton
field during inflation can be written as follows
\begin{equation}
u_k=u(\tau,\textbf{k})=\frac{i H}{\sqrt{4\epsilon
c_s k^3}}(C_{+}(1+i k c_s\tau
)e^{-i k c_s\tau}+C_{-}(1-i k c_s\tau)e^{i k c_s\tau})
\end{equation}
In the standard Bunch-Davies vacuum we have $C_{+}=1$ and
$C_{-}=0$. Now we allow a small deviation from the
Bunch-Davies vacuum by turning on a small finite number $C_{-}$,
and calculate the corrections to the shape of non-Gaussianities we
found in the Bunch-Davies vacuum. For simplicity we only consider
the corrections to the leading order non-Gaussianities
$A_{\lambda}$, $A_{c}$ in the small sound speed $c_s\ll 1$ limit 
(the corrections to the other
shapes of non-Gaussianities due to a non-standard choice of vacuum
can be worked out by a similar procedure).
The computations of the three point functions of the primordial
perturbations essentially go through as before. The first
sub-leading correction to Bunch-Davies vacuum result is simply to
replace one of the three $u(\tau,\textbf{k})$'s with its $C_{-}$
component. 
(A correction of $\CO(C_-)$ in $u(0,\bk)$ gives a term which has
the same shape as in the Bunch-Davies vacuum case. We do not include
them here.)
This does not change the common factor $\frac{1}{k_1^3
k_2^3 k_3^3}$ but will simply replace one of the $k_i$ in the
shapes $A_{\lambda}$, $A_c$  with $-k_i$. We denote the
corresponding corrections $\tilde{A}_{\lambda}$, $\tilde{A}_{c}$.
We immediately find the corrections as
\begin{eqnarray}
\tilde{A}_{\lambda} &=& Re(C_{-})
(\frac{1}{c_s^2}-1-\frac{2\lambda}{\Sigma})\frac{3k_1^2k_2^2k_3^2}{2}
\nonumber \\ &\times&
(\frac{1}{(k_1+k_2-k_3)^3}+\frac{1}{(k_1-k_2+k_3)^3}+\frac{1}{(-k_1+k_2+k_3)^3}),
\\
\tilde{A}_c &=& Re(C_{-}) (\frac{1}{c_s^2}-1)\sum_{p=1 }^3
(-\frac{1}{K}\sum_{i>j}k_i^2k_j^2+\frac{1}{2K^2}\sum_{i\neq
j}k_i^2k_j^3+\frac{1}{8}\sum_{i}k_i^3)|_{k_p\rightarrow -k_p} .
\end{eqnarray}


We can estimate the size of the non-Gaussianities
$\tilde{A}_{\lambda}$ and $\tilde{A}_{c}$ according to the WMAP
ansatz. This estimate is usually done in the equilateral triangle limit; 
we find
\begin{eqnarray}
\tilde{f}_{NL}^{\lambda} &=&  -5 Re(C_{-})
(\frac{1}{c_s^2}-1-\frac{2\lambda}{\Sigma}) ~, \nonumber \\
\tilde{f}_{NL}^c &=& \frac{25}{4} Re(C_{-}) (\frac{1}{c_s^2}-1) ~.
\end{eqnarray}

\begin{figure}
\begin{center}
\epsfig{file=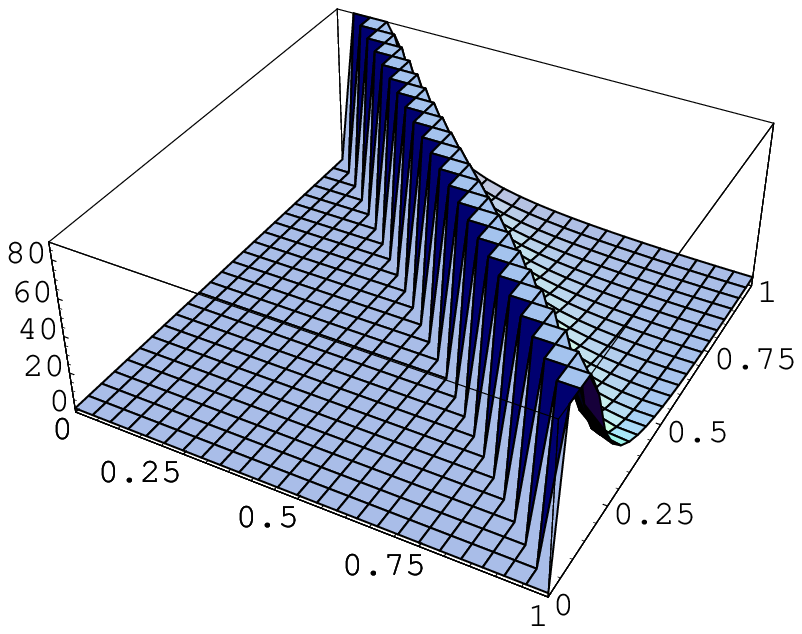, width=9cm}
\end{center}
\medskip
\caption{The shape of $|\tilde\CA_\lambda|/k_1k_2k_3$} 
\label{BD1}
\end{figure}

\begin{figure}
\begin{center}
\epsfig{file=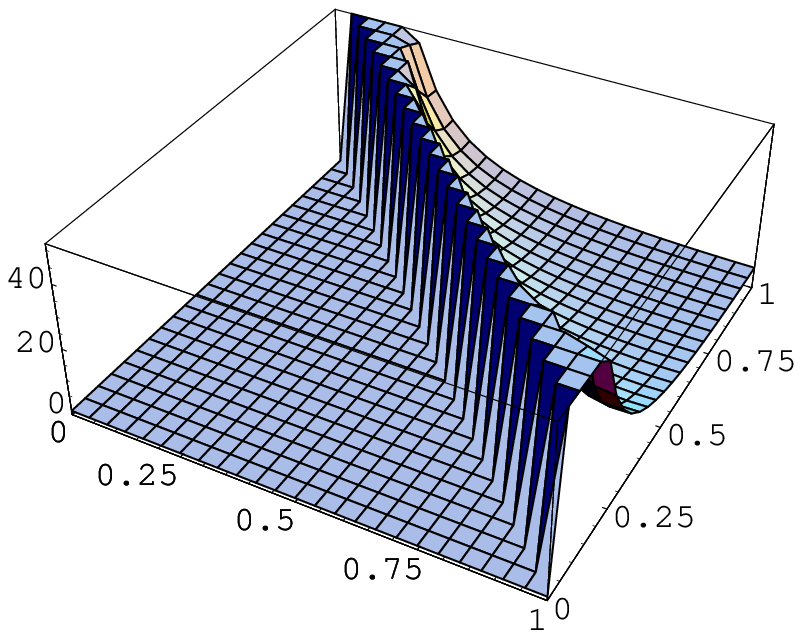, width=9cm}
\end{center}
\medskip
\caption{The shape of $|\tilde\CA_c|/k_1k_2k_3$} 
\label{BD2}
\end{figure}

If the sound speed $c_s$ is sufficiently small,
the effects of a slight deviation from the Bunch-Davies vacuum is 
potentially observable by future experiments. We plot the shapes
of the non-Gaussianities $\tilde{A}_{\lambda}$ and $\tilde{A}_{c}$
in Fig. \ref{BD1} and Fig. \ref{BD2}. We see the shapes of these
corrections are very distinctive and in fact dramatically different from that of the
DBI inflation or slow roll inflation. In particular, these shapes
are highly peaked at the ``folded triangle'' limit where
$k_3 \approx k_1+k_2$ for arbitrary values of $k_1$ and $k_2$. This
feature is not shared by other known sources
of non-Gaussianities, and so measurements of the shape of non-Gaussianities
could in principle be an excellent
probe of the choice of inflationary vacuum.

Note that, while the rising behavior of the non-Gaussianity 
in the folded triangle limit is the signal of the non-Bunch-Davies
vacuum, the divergence at the limit e.g.~$k_1+k_2-k_3=0$ 
is artificial. This divergence is present
because we have assumed that such a non-standard vacuum existed in
the infinite
past. Realistically there should be a cutoff at a large momentum
$M$ for $k/a$, where $k$ is a typical value of $k_{1,2,3}$. This
amounts to a cutoff for $\tau$ at $\tau_c = - M/Hk$. Since the
integrand is regulated at $\tau= -1/Kc_s$ due to its rapid
oscillation, if $\tau_c < -1/Kc_s$, the cutoff $M$ has no effects to
our calculation. That is, for $K \gg kH/M c_s$, we will see the behaviors
shown in Fig.~\ref{BD1} \& \ref{BD2} near the folded triangle limit. 
But within  $K <k H/M c_s$, the
cutoff takes effect first, the divergence behavior will be
replaced. The details depend on the nature of the cutoff, e.g. a
naive sharp cutoff will introduce oscillatory behavior.

\section{Conclusion}

The forthcoming suite of cosmological experiments will nail down
with ever greater precision the parameters
of the inflationary model that yielded our
homogeneous, isotropic universe.
Some measurements, like the value of the spectral index and the nature
of its running, are guaranteed to occur.
Others, like a detection of primordial gravitational waves,
are not necessarily expected to occur on theoretical grounds (since models
with very small $r$ seem more natural as quantum field theories), but would
be tremendously exciting and instructive if they do.
The discovery of significant non-Gaussian scalar fluctuations
falls into this latter category.  While the simplest models of inflation
do not produce this phenomenon, its discovery would tell us something
qualitatively important about the inflationary epoch, and experiments
sensitive enough to measure $\vert f_{NL} \vert  \geq 5$ 
will be launched in the next two years.
For this reason, we feel it is worthwhile to parametrize the
reasonable possibilities, and understand the qualitative physics of
the models that produce them.

In this paper, we have taken some steps in this direction for generic
single-field models.
There are several clear directions for further work:

\medskip
\noindent
$\bullet$
It would be nice
to derive the same formulae governing non-Gaussianities as arising
directly from symmetry principles.  Perhaps these would be
encapsulated most neatly in
a hypothetical dual, non-gravitational theory.
For the models with $c_s << 1$, this theory may have novel properties.

\medskip
\noindent
$\bullet$
Higher derivative terms play a significant role in the dynamics of
those single-field models which produce striking non-Gaussian
signatures.  One class of models where such terms are important, the
DBI inflation
\cite{Silverstein:2003hf,Alishahiha:2004eh,Chen:2004gc,Chen:2005ad},
has a reasonable microscopic justification in string theory.  It would
be interesting to find other
examples where one can microphysically justify the study of dynamics
that is very sensitive to higher derivative terms.

\medskip
\noindent
$\bullet$
We have focused here on single-field models.  It is a logical
possibility that our 60 e-foldings arose from a multi-field inflationary
model.  This could be motivated if, for instance, $r$ is measured
to be non-negligible.  In slow-roll models, measurable $r$ implies
an inflaton that traversed a super-Planckian distance in field space
\cite{Lythbound}, as in chaotic inflation \cite{chaotic}.
At least in string theory, this is difficult to accommodate in single-field
models \cite{Dine}, but could conceivably happen in a multi-field
setting \cite{Nflation}.
For this and other reasons, it would be worthwhile to develop a general
framework for analyzing non-Gaussianities in multi-field models.
Examples of multi-field models with significant non-Gaussianity appear
in \cite{LindeMuk,Bernardeau,Enqvist:2004ey}.
Formalisms to compute the non-Gaussianities in large classes of such models
are developed in \cite{Rigo,SL,LythNG,Wands}.

\medskip
\section*{Acknowledgments}

We thank D. Chung, G. Geshnizjani, A. Giryavets, L. Hui, S. Kecskemeti,
X. Liu, J. Maiden, 
A. Maloney, E. Silverstein, B. Underwood, L. Verde and E. Weinberg
for discussions. 
We thank Paulo Creminelli, Juan Maldacena, Leonardo Senatore, Matias
Zaldarriaga and the authors of Ref.~\cite{Cheung:2007sv} 
for many discussions and comments on the consistency
condition and non-BD vacuum, and for providing us the motivation to check
and correct the numerical coefficients in Appendix \ref{AppCorr}.
XC thanks the physics department of
the University of Wisconsin at Madison for hospitality.
The work of XC was supported in part by the U.S. Department of Energy.
The work of MH and GS
was supported in part by NSF CAREER Award No. PHY-0348093,
DOE grant DE-FG-02-95ER40896, a Research Innovation Award and a Cottrell
Scholar Award from Research Corporation.
The work of SK was supported by a David and Lucile Packard Foundation
Fellowship for Science and Engineering, the DOE under contract
DE-AC02-76SF00515, and the NSF under grant 0244728.

\appendix

\section{Corrections to $u_k$}
\label{SecukCorr}

In this appendix, we calculate the $\CO(\epsilon)$ correction to the
solution of Eq.~(\ref{quadeom}), generalizing the method of 
\cite{Stewart:1993bc,Easther:2001fi}.

We define
\bea
y \equiv \frac{c_s k}{a H}
\eea
and write the equation of motion of the quadratic action
\bea
v_k'' + c_s^2 k^2 v_k - \frac{z''}{z} v_k =0
\label{eomquard}
\eea
in terms of $y$. Note that generally $c_s$ is a (slowly varying)
function of time.
Using 
\bea
\frac{z''}{z} = 2a^2 H^2 (1-\half \epsilon + \frac{3}{4} \eta -
\frac{3}{2} s) + \CO(\epsilon^2)
\eea
we get
\bea
(1-2\epsilon -2s) ~y^2~ \frac{d^2 v_k}{dy^2} -s y~\frac{d v_k}{dy} +
y^2 v_k - ( 2-\epsilon+\frac{3}{2}\eta -3s) v_k =0 ~.
\eea
The solution of this differential equation is given by 
\bea
v_k = y^{\half (1+s)} \left[ C_1 H_\nu^{(1)} \left
( (1+\epsilon+s)y\right) + C_2 H_\nu^{(2)}\left((1+\epsilon+s)y\right)
\right] ~,
\label{vksolcorr}
\eea
where
\bea
\nu = \frac{3}{2} + \epsilon + \frac{\eta}{2} +\frac{s}{2} ~.
\eea
The Bunch-Davies (BD) vacuum corresponds to $C_2=0$. To determine the coefficient $C_1$, we need to look at
the large $k$ behavior of the equation (\ref{eomquard}),
\bea
v_k'' + c_s^2 k^2 v_k =0 ~.
\label{eomquardlargek}
\eea
The behavior is more general than the usual case where the sound speed
is constant, because here we allow $c_s$ to vary slowly. Using a
similar approach and defining $\ty \equiv -c_s k \tau = (1+\epsilon
+\CO(\epsilon^2)) y$, we get the
solution for (\ref{eomquardlargek}) with positive energy (BD vacuum), 
\bea
v_k \to \frac{1}{\sqrt{2 c_s k}} e^{i[(1+s)\ty - \frac{\pi}{4} s ]}
~,
\label{vklargek}
\eea
up to a constant phase. Here the coefficient is determined by the
quantization condition (Wronskian condition), $v_k^*
\frac{dv_k}{d\tau} - v_k \frac{dv_k^*}{d\tau} = -i$, to first order
$\CO(\epsilon)$. 
Notice that in (\ref{vklargek}), the sound speed $c_s$ runs as a
function of $y$. So we can
expand it as
\bea
v_k \to \frac{1}{\sqrt{2 c_{s0} k}} \left(\frac{y}{y_0}\right)^{s/2} 
e^{i[(1+s)\ty - \frac{\pi}{4} s ]} ~,
\eea
where the subscript $0$ on $c_s$ denotes the evaluation at $y_0$.
Expanding (\ref{vksolcorr}) in the same limit,
\bea
v_k \to C_1 \sqrt{\frac{2}{\pi}}~ \frac{1}{\sqrt{1+\epsilon+s}}~
y^{s/2}~
e^{i[y(1+\epsilon+s)-\frac{\pi}{2}\nu-\frac{\pi}{4}]} ~, ~~~~ y\gg 1,
\eea
we find
\bea
C_1 = - \frac{\sqrt{\pi}}{2} \frac{1}{\sqrt{c_{s k} k}} 
(1+\frac{\epsilon}{2} + \frac{s}{2})
~e^{i\frac{\pi}{2} (\epsilon + \frac{\eta}{2}) } ~.
\eea

Note a convention for the variables used here: the variables such as
$c_{sk}$ with the subscript $k$ are evaluated at
\bea
y_0=\frac{k c_{s k}}{a_k H_k}=1 ~;
\eea
in some later formulae, the variables without
the subscript such as $c_s$ mean that they are functions of $y$.

From the definition 
\bea
v_k\equiv z u_k ~, ~~~~z \equiv \frac{a \sqrt{2\epsilon}}{c_s} ~,
\eea
we get the expression for $u_k$ to order $\CO(\epsilon)$,
\bea
u_k(y) = -\frac{\sqrt{\pi}}{2\sqrt{2}} ~\frac{H}{\sqrt{\epsilon c_s}}~
\frac{1}{k^{3/2}} (1+\frac{\epsilon}{2} +\frac{s}{2} )
~e^{i \frac{\pi}{2} (\epsilon + \frac{\eta}{2}) }~ y^{3/2}
H_\nu^{(1)} \left( (1+\epsilon+s)y \right) ~.
\label{ukepsilon}
\eea

As an application, we use the $y \to 0$ limit of Eq.~(\ref{ukepsilon})
to derive an expression for the density perturbation to order
$\CO(\epsilon)$. To do this, we use the expansion of the Hankel
function in the $y\ll 1$ limit,
\bea
H_\nu^{(1)}(y) \to -i \frac{1}{\sin \nu \pi} \frac{1}{\Gamma(-\nu+1)}
\left(\frac{y}{2}\right)^{-\nu} ~,
\eea
and get
\bea
u_k(0) = \frac{i H_k}{2\sqrt{c_{s k}\epsilon_k}} \frac{1}{k^{3/2}}
\left(1-(\bc_2+1)\epsilon - \frac{\bc_2}{2}\eta -(\frac{\bc_2}{2} +1)s
\right) e^{i\frac{\pi}{2} (\epsilon + \frac{\eta}{2}) } ~,
\label{uk0corr}
\eea
where 
$$\bc_2 \equiv \bc_1-2+\ln 2 \approx -0.73 $$
and $\bc_1 = 0.577\cdots$ is the Euler constant.
Hence the density perturbation is
\bea
\sqrt{P^{\zeta}_k} &=& \sqrt{\frac{k^3}{2\pi^2}}~ |u_k(y=0)| 
\nonumber \\
&=& \frac{1}{\sqrt{8\pi^2}} \frac{H_k}{\sqrt{c_{s k} \epsilon_k}} 
\left(1-(\bc_2+1)\epsilon - \frac{\bc_2}{2}\eta -(\frac{\bc_2}{2} +1)s
\right) +\CO(\epsilon^2) ~.
\label{Pk}
\eea
This generalizes the result (31) of
Ref.~\cite{Stewart:1993bc} to the case with 
running sound speed.

\section{Details on the correction terms}
\label{AppCorr}
In this Appendix, we provide details on the correction terms in
Sec.~\ref{SecCorr}.
Let us look at the first integration (\ref{IntFirst}) in
Sec.~\ref{SecCubic},
\bea
-6i \int d\tau ~a~ f_1(\tau) \prod_i u(0,\bk_i) 
\frac{d}{d\tau} u^*(\tau,\bk_i)
\cdot (2\pi)^3 \delta^3(\sum_i \bk_i) + {\rm c.c.} ~,
\label{IntFirst1}
\eea
where 
\bea
f_1(\tau) = \frac{\epsilon}{H c_s^4} (c_s^2-1) + \frac{2\lambda}{H^3} ~.
\eea
We have evaluated the leading contribution of this integral in
Sec.~\ref{SecCubic}. The
corrections come from several different places. 

The first is from the time variation in $f_1(\tau)$,
\bea
f_1(\tau) &=& f_1(t_K) + \frac{\partial f_1}{\partial t} (t-t_K) +
\CO (\epsilon^2 f_1) \nonumber \\
&=& f_1(\tau_K) - \frac{\partial f_1}{\partial t} \frac{1}{H_K} \ln
\frac{\tau}{\tau_K} + \CO(\epsilon^2 f_1)
~.
\label{fcorr1}
\eea
We choose to eventually evaluate all the variables at the time 
$\tau_K$, which is defined as the moment when the wave-number 
$K=k_1+k_2+k_3$
exits the horizon $K c_{sK} = a_K H_K$, at which
\bea
\tau_K \equiv -\frac{1}{K c_{s K}} + \CO(\epsilon) ~.
\label{tau0}
\eea
All the subscripts $K$ denote the evaluation at the horizon exit
point defined in (\ref{tau0}).
The $\partial f_1/\partial t$ can be expressed in terms of the slow
variation parameters
\bea
\frac{\partial f_1}{\partial t} = 
(\eta \epsilon + \epsilon^2) (c_s^{-2} - c_s^{-4}) + \epsilon s
(-2c_s^{-2} +4c_s^{-4}) + 2 l \lambda H^{-2} + 6 \epsilon \lambda
H^{-2} ~,
\eea
where we have defined 
\bea
\l \equiv \frac{\dot \lambda}{\lambda H} ~.
\eea
We assume that the time variation of $\lambda$ is slow,
$l=\CO(\epsilon)$.
Plugging the corrections terms of (\ref{fcorr1}) into
(\ref{IntFirst1}) and evaluating the rest in leading orders, we
get\footnote{The integration 
$\int_{-\infty}^0 dx \ln(-x)~ e^{ix} = i\bc_1 - \frac{\pi}{2}$ has been
used. Similar types of integrations will be used later.}
\bea
\Delta \CA &\supset& 
\frac{9}{4} (1-\frac{2}{3} \bc_1)
\left( (-\epsilon-\eta)(\frac{1}{c_s^2}-1) +
s(\frac{4}{c_s^2} -2) + (2l + 6\epsilon) \frac{\lambda}{\Sigma}
\right)_K \frac{k_1^2 k_2^2 k_3^2}{K^3} ~.
\label{DA11}
\eea

The second comes from the correction to the scale factor $a \approx
-\frac{1}{H\tau}$. This can be obtained from the relation
\bea
d\tau = \frac{dt}{a} = - \frac{1}{H} d\left( \frac{1}{a} \right) 
\label{taua}
\eea
and the expansion
\bea
\frac{1}{H} = \frac{1}{H_K} + \epsilon (t-t_K) +\CO(\epsilon^2)~.
\eea
Integrating (\ref{taua}) we can get the following expansion
\bea
a = -\frac{1}{H_K \tau} - \frac{\epsilon}{H_K \tau} 
+ \frac{\epsilon}{H_K \tau} 
\ln (\tau/\tau_K) + \CO(\epsilon^2) ~.
\label{Expansiona}
\eea
Plugging this correction term into (\ref{IntFirst1}), we obtain
\bea
\Delta \CA \supset (\bc_1 -\frac{1}{2})~ \epsilon 
\left( \frac{1}{c_s^2} -1 -
\frac{2\lambda}{\Sigma} \right)_K \frac{3k_1^2 k_2^2 k_3^2}{2K^3} ~.
\label{DA12}
\eea

The third comes from the correction term to $u(\tau, \bk_i)$ which we
obtained in Sec.~\ref{SecukCorr}. We first look at the corrections to
the factor $u(0,\bk_i)$ in (\ref{IntFirst1}). The corrections to the
final result come not only from the corrections in the bracket of
(\ref{uk0corr}), but also from the running from $k_i$ to $K$
\bea
\frac{H_{k_i}}{\sqrt{c_{s i} \epsilon_i}} = 
\frac{H_{K}}{\sqrt{c_{s K} \epsilon_K} }
\left( 1-(\epsilon + \frac{\eta}{2} + \frac{s}{2}) \ln \frac{k_i}{K}
\right) +\CO(\epsilon^2) ~.
\label{ExpansionH}
\eea
So these add a correction term to the three-point correlation
function\footnote{To make sure that the expansion such as
(\ref{ExpansionH})
is perturbative, we need $k_i \gg \CO(K e^{-1/\epsilon})$. Note that
this condition still allows one of the momenta to be much smaller
than the others, e.g.~$(k_3/k_1)^2 \ll \epsilon$.

In order for the expansion such as (\ref{Expansiona}) to be
perturbative, we need $\tau_K e^{-1/\epsilon} \gg \tau \gg \tau_K
e^{1/\epsilon}$. So it appears that the integration over $\tau$ can
only be taken from
$\tau_K e^{1/\epsilon}$ to $\tau_K e^{-1/\epsilon}$. 
We first look at the upper bound.
Since the mode
$k_i$ exits the horizon at $\tau_i \approx -1/k_i c_{si}$. At the upper
bound of $\tau$, all modes have exited the horizon and their
amplitudes are frozen. So the error introduced by including the
integration from $\tau_K e^{-1/\epsilon}$ to $0$ is of order
$\CO(\frac{k_ic_{si}} {K C_{sK}} e^{-1/\epsilon}) \sim \CO
(e^{-1/\epsilon})$.
We next look at the lower bound. For the range of $k_i$ that we are
interested in, at $\tau \sim \tau_K e^{1/\epsilon}$ all modes are well
within the horizon. Their contributions are regulated away due to
their rapid oscillation. 
Therefore, to order $\CO(e^{-1/\epsilon})$, we can effectively take the
integration range for $\tau$ from $-\infty$ to 0. }

\bea
\Delta \CA &\supset& 
\left( -3(\bc_2+1)\epsilon - \frac{3\bc_2}{2}\eta
-3(\frac{\bc_2}{2}+1)s 
- (\epsilon + \frac{\eta}{2} +\frac{s}{2}) \ln \frac{k_1k_2k_3}{K^3}
\right) \nonumber \\
&\times& \left(\frac{1}{c_s^2}-1
-\frac{2\lambda}{\Sigma} \right)_K \frac{3k_1^2k_2^2k_3^2}{2K^3} ~.
\label{DA13}
\eea

We next look at the corrections to the factor $\frac{d}{d\tau}
u^*(\tau,\bk_i)$ in (\ref{IntFirst1}). To do this we use
(\ref{ukepsilon}) and expand the pre-factor around $k=K$,
\bea
u_k(y) &=& 
-\frac{\sqrt{\pi}}{2\sqrt{2}} ~\frac{H_K}{\sqrt{\epsilon_K c_{s K}}}~
\frac{1}{k^{3/2}} \left(1+\frac{\epsilon}{2} +\frac{s}{2}
+ (\epsilon + \frac{\eta}{2} + \frac{s}{2}) \ln \frac{\tau}{\tau_K} 
\right)
~e^{i\frac{\pi}{2} (\epsilon +\eta)} \nonumber \\ 
&\times& y^{3/2}
H_\nu^{(1)} \left( (1+\epsilon+s)y \right) ~.
\label{ukepsilon2}
\eea
Denoting $\Delta u(\tau,k_i)$ as the corrections to the leading order,
we have
\bea
\Delta u^*(\tau,k_i) &=& -\frac{1}{2} 
\frac{H_K}{\sqrt{c_{s K}\epsilon_K}} \frac{1}{k_i^{3/2}}
e^{-i\frac{\pi}{2}(\epsilon+\eta)} e^{-ix} 
\nonumber \\
&\times& [-i(\epsilon+s) + (\epsilon + s) x + i s x^2 
\nonumber \\
&+& 
(i(\epsilon +\frac{\eta}{2} +\frac{s}{2}) - (\epsilon +\frac{\eta}{2}
+\frac{s}{2}) x - i x^2 s) \ln\frac{\tau}{\tau_K}
\nonumber \\
&+& \sqrt{\frac{\pi}{2}} e^{ix} (\epsilon +\frac{\eta}{2}
+\frac{s}{2}) x^{3/2} \frac{d H_\nu^*}{d\nu} ] ~,
\label{Du*}
\eea
where $x\equiv -k_i c_{s K} \tau$. The above corrections include
those in the first line of (\ref{ukepsilon2}), in $y^{3/2}$ where
$y=-k_i c_{s K} \tau (1-\epsilon - s \ln \frac{\tau}{\tau_K})
+\CO(\epsilon^2)$
is used, and in $H^{(1)*}_\nu((1+\epsilon+s)y)$ 
which includes corrections in the index $\nu$ and corrections in the
variable $y$.
Differentiate (\ref{Du*}), we have
\bea
\frac{d}{d\tau} \Delta u^*(\tau,k_i) &=& 
\frac{1}{2} \frac{H_K}{\sqrt{c_{s K}\epsilon_K}} \frac{1}{k_i^{3/2}}
e^{-i\frac{\pi}{2}(\epsilon+\eta)} k_i c_{s K} e^{-ix} 
\nonumber \\
&\times& [ -(\epsilon +\frac{\eta}{2} +\frac{s}{2}) + (\epsilon
+\frac{\eta}{2} +\frac{s}{2}) \frac{i}{x} - i\epsilon x + s x^2 
\nonumber \\
&+& ( i\epsilon +\frac{i}{2} \eta - \frac{3}{2}i s - s x) x
\ln\frac{\tau}{\tau_K} 
\nonumber \\
&+& \frac{\sqrt{\pi}}{\sqrt{2}} e^{ix} 
\frac{d}{dx} (x^{3/2} \frac{d H^*_\nu}{d\nu}) 
(\epsilon +\frac{\eta}{2} +\frac{s}{2})] ~.
\label{dDu*dtau}
\eea
The first two lines in the square bracket in (\ref{dDu*dtau}) contribute
\bea
\Delta \CA &\supset& \frac{3}{4}
\left( \frac{1}{c_s^2}-1-\frac{2\lambda}{\Sigma} \right)_K
\Bigg( ((3-6\bc_1)\epsilon + (\frac{9}{2}-3\bc_1)\eta +
(3\bc_1-\frac{17}{2})s) \frac{k_1^2 k_2^2 k_3^2}{K^3}
\nonumber \\
&+& (\epsilon +\frac{\eta}{2} +\frac{s}{2}) 
(\frac{1}{K^2} \sum_{i\neq j} k_i^2 k_j^3 -
\frac{2}{K} \sum_{i>j} k_i^2 k_j^2 ) \Bigg) ~.
\label{DA14}
\eea
The last term in (\ref{dDu*dtau}) involves special functions and
contributes
\bea
\Delta \CA \supset  \frac{3}{4} 
(\epsilon + \frac{\eta}{2} + \frac{s}{2}) 
\left( \frac{1}{c_s^2}-1-\frac{2\lambda}{\Sigma} \right)_K
R_1(k_1,k_2,k_3) + {\rm sym} ~,
\label{DA15}
\eea
where
\bea
&&R_1(k_1,k_2,k_3) = \frac{k_2^2 k_3^2}{k_1}
{\rm Re}\left[ \int_0^{\infty} dx_1~ h^*(x_1)
( 1- i\frac{k_2+k_3}{k_1} x_1) e^{-i\frac{k_2+k_3}{k_1}x_1} \right] ~,
\\
&& x_1\equiv -k_1 c_{sK} \tau ~,
\nonumber \\
&&h(x) = - 2i e^{ix} + i e^{-ix} (1+ix) [\Ci(2x)+i\Si(2x)]
-i\pi \sin x + i\pi x \cos x ~.
\eea
We have used the following relations,
\bea
&& x H^{(1)}_{\nu-2}(x) + x H^{(1)}_\nu =
2(\nu-1) H^{(1)}_{\nu-1}(x) ~,
\\
&&\left[ \frac{\partial J_\nu(x)}{\partial \nu} \right]_{\nu=\half}
= \left( \half \pi x \right)^{-\half}
[\sin x \Ci(2x) -\cos x \Si(2x)] ~,
\\
&&\left[ \frac{\partial N_\nu(x)}{\partial \nu} \right]_{\nu=\half}
= \left( \half \pi x \right)^{-\half}
\{\cos x \Ci(2x) +\sin x [\Si(2x)-\pi]\} ~,
\\
&&\left[ \frac{\partial J_\nu(x)}{\partial \nu} \right]_{\nu=-\half}
= \left( \half \pi x \right)^{-\half}
[\cos x \Ci(2x) + \sin x \Si(2x)] ~,
\\
&&\left[ \frac{\partial N_\nu(x)}{\partial \nu} \right]_{\nu=-\half}
= - \left( \half \pi x \right)^{-\half}
\{\sin x \Ci(2x) -\cos x [\Si(2x)-\pi]\} ~.
\eea

The same procedure can be repeated for the second integration
(\ref{IntSecond}) in
Sec.~\ref{SecCubic},
\bea
&&-2i \int_{-\infty}^0 d\tau ~a^2 ~f_2(\tau)~ ( u(0,\bk_1) u(0,\bk_2)
u(0,\bk_3) \nonumber \\ 
&\times& u^*(\tau,\bk_1) \frac{du^*(\tau,\bk_2)}{d\tau}
\frac{du^*(\tau,\bk_3)}{d\tau} + {\rm sym} ) ~ 
(2\pi)^3 \delta^3(\sum_i\bk_i) + {\rm c.c.} ~,
\label{IntSecond2}
\eea
where
\bea
f_2 = \frac{\epsilon}{c_s^4} (3- 3c_s^2 -\epsilon) ~.
\eea
 From the variation of $f_2$, we get the correction term
\bea
\Delta \CA \supset
\left( (\frac{3\eta}{4}-3s)(\frac{1}{c_s^2}-1)
-\frac{3s}{2} \right) 
\left( (1-2\bc_1) \frac{1}{K}\sum_{i>j} k_i^2 k_j^2
- (1-\bc_1) \frac{1}{K^2} \sum_{i\ne j} k_i^2 k_j^3 \right) ~. 
\nonumber \\
\label{A21}
\eea
 From the correction term to the scale factor $a$, we get
\bea
\Delta \CA \supset \epsilon \left(\frac{1}{c_s^2}-1 \right)_K
\left( -(\frac{3}{2}+3\bc_1) \frac{1}{K}\sum_{i>j} k_i^2 k_j^2
+ \frac{3\bc_1}{2} \frac{1}{K^2} \sum_{i\ne j} k_i^2 k_j^3 \right)
~.
\label{A22}
\eea
 From the correction term to $u(0, k_i)$, we get
\bea
\Delta \CA &\supset& 
\frac{3}{4} \Bigg( 3(\bc_2+1) \epsilon + \frac{3\bc_2}{2} \eta +
3(\frac{\bc_2}{2}+1) s + (\epsilon +\frac{\eta}{2} +\frac{s}{2}) 
\ln\frac{k_1 k_2 k_3}{K^3} \Bigg)
\nonumber \\
&\times& \left(\frac{1}{c_s^2}-1 \right)_K
\left( \frac{2}{K} \sum_{i>j} k_i^2 k_j^2 - \frac{1}{K^2} \sum_{i\neq
j} k_i^2 k_j^3 \right) ~.
\label{A23}
\eea
The correction to $u^*(\tau,k_i)$ in (\ref{Du*}) contributes
\bea
\Delta \CA &\supset& -\frac{3}{4} \left(\frac{1}{c_s^2}-1\right)_K 
\Bigg( (3-6\bc_1) s \frac{k_1^2 k_2^2 k_3^2}{K^3}
\nonumber \\
&+& (-(1+2\bc_1)\epsilon + (\half-\bc_1) \eta -(\frac{3}{2}+\bc_1)s)
\frac{1}{K} \sum_{i>j} k_i^2 k_j^2
\nonumber \\
&+& (\bc_1\epsilon -\half(1-\bc_1)\eta + \half (1+\bc_1)s) 
\frac{1}{K^2} \sum_{i\neq j} k_i^2 k_j^3 \Bigg) 
\label{A24}
\eea
and 
\bea
\Delta A \supset -\frac{3}{4} \left(\frac{1}{c_s^2}-1\right)_K 
\left(\epsilon +\frac{\eta}{2} +\frac{s}{2} \right) 
\left( \frac{k_2^2 k_3^2}{k_1} G_1 + {\rm sym} \right) ~,
\label{A25}
\eea
where
\bea
G_1 \equiv {\rm Re} 
\left[ \int_0^{\infty} dx_1 h^*(x_1) e^{-i\frac{k_2+k_3}{k_1}x_1} \right] ~.
\eea
The correction to $\frac{d}{d\tau} u^*(\tau,k_i)$ in (\ref{dDu*dtau}) 
contributes
\bea
\Delta \CA &\supset& \frac{3}{4} \left(\frac{1}{c_s^2}-1\right)_K 
\Bigg( ( (2+4\bc_1)\epsilon + (-1+2\bc_1) \eta + (3-6\bc_1)s ) 
\frac{1}{K} \sum_{i>j} k_i^2 k_j^2
\nonumber \\
&+& ( -2\bc_1\epsilon +(1-\bc_1)\eta +(-4+6\bc_1)s)
\frac{1}{K^2} \sum_{i\neq j} k_i^2 k_j^3
\nonumber \\
&+& (2\epsilon + \eta +s) k_1 k_2 k_3
\nonumber \\
&+& (1-2\bc_1)s (\frac{1}{K^3} \sum_{i\neq j} k_i^2 k_j^4
+\frac{2}{K^3} \sum_{i> j} k_i^3 k_j^3)
\nonumber \\
&-& (2\epsilon +\eta + s) 
\left(\sum_i k_i^3 + \sum_{i\neq j} k_i k_j^2 + \sum_i k_i^3
~{\rm Re} \int_0^{\infty} dx_K \frac{e^{-ix_K}}{x_K} \right) 
\Bigg)
\label{A26} \\
&&  ( x_K \equiv -K c_{sK} \tau )
\nonumber
\eea
and
\bea
\Delta \CA \supset 
\frac{3}{4} \left(\frac{1}{c_s^2}-1\right)_K 
\left(\epsilon +\frac{\eta}{2} +
\frac{s}{2}\right) ( \tilde M_1 + {\rm sym}) ~,
\label{A27}
\eea
where 
\bea
\tilde M_1 \equiv - k_1 {\rm Re} \int_0^{\infty} dx_1 \frac{1}{x_1} 
(k_2^2 + k_3^2 +i k_2 k_3 \frac{k_2+k_3}{k_1} x_1)
e^{-i\frac{k_2+k_3}{k_1} x_1} \frac{d h^*(x_1)}{dx_1} ~.
\label{tildeM}
\eea
Notice that in (\ref{A26}), the last term is divergent. This
divergence is cancelled by the divergence that appears in
(\ref{tildeM}) using the limit $h(x) \to (-2+\bc_1)i +i \ln 2x +
\CO(x^2)$ as $x \to 0$. So we can re-define 
\bea
M_1 +{\rm sym} \equiv \tilde M_1 - {\rm Re} (k_2^3 + k_3^3) 
\int_0^{\infty} dx_K \frac{e^{-ix_K}}{x_K} + {\rm sym}
\eea
to absorb this divergence.

The following are the corrections to the third integration (\ref{IntThird}) 
\bea
&&-2i \int_{-\infty}^{0} d\tau ~a^2~ f_3(\tau)~ 
(~ u(0,\bk_1) u(0,\bk_2)
u(0,\bk_3) \nonumber \\
&\times& u^*(\tau,\bk_1) u^*(\tau,\bk_2) u^*(\tau,\bk_3)
(-\bk_2\cdot\bk_3) + {\rm sym} ) \cdot (2\pi)^3 \delta^3(\sum_i \bk_i)
+ {\rm c.c.}
\eea
with
\bea
f_3 = -\frac{\epsilon}{c_s^2} (1-c_s^2 -2s +\epsilon) ~.
\eea
From the variation of $f_3$, we get
\bea
\Delta \CA &\supset&
\left( (\frac{\eta}{4} -\frac{s}{2})
(1-\frac{1}{c_s^2}) 
+ \frac{s}{2}
\right)_K \nonumber \\
&\times& \left( \frac{1-\bc_1}{2} \sum_i k_i^3 - \half k_1 k_2 k_3
+\half \sum_{i\neq j}k_i k_j^2 
+ \frac{1-2\bc_1}{K} \sum_{i>j} k_i^2 k_j^2 - 
\frac{1-\bc_1}{K^2} \sum_{i\neq j} k_i^2 k_j^3 \right)~. \nonumber\\ 
\label{A31}
\eea
From the corrections to $a$, we get
\bea
\Delta \CA
&\supset& \frac{\epsilon}{2} 
\left( 1-\frac{1}{c_s^2} \right)_K \nonumber \\
&\times& \left( \frac{-\bc_1}{2} \sum_i k_i^3 - \half k_1 k_2 k_3
+\half \sum_{i\neq j}k_i k_j^2 
- \frac{1+2\bc_1}{K} \sum_{i>j} k_i^2 k_j^2 
+\frac{\bc_1}{K^2} \sum_{i\neq j} k_i^2 k_j^3 \right)~. 
\nonumber\\
\label{A32}
\eea
From the corrections to $u(0,k_i)$, we get
\bea
\Delta \CA &\supset& 
-\frac{1}{4} \Bigg( 3(\bc_2+1) \epsilon + \frac{3\bc_2}{2} \eta +
3(\frac{\bc_2}{2}+1) s + (\epsilon +\frac{\eta}{2} +\frac{s}{2}) 
\ln\frac{k_1 k_2 k_3}{K^3} \Bigg)
\nonumber \\
&\times& \left(\frac{1}{c_s^2}-1 \right)_K 
\left( \half \sum_i k_i^3 + \frac{2}{K} \sum_{i>j} k_i^2 k_j^2 -
\frac{1}{K^2} \sum_{i\neq j} k_i^2 k_j^3 \right) ~.
\label{A33}
\eea
Corrections from $u^*(\tau,k_i)$ give
\bea
\Delta \CA &\supset&
-\frac{1}{8} \left(\frac{1}{c_s^2}-1\right)_K
\nonumber \\
&\times& \Bigg( (3\bc_1 \epsilon + \frac{3}{2}(-1 +\bc_1) \eta +
\frac{3}{2}(1+\bc_1) s ) \sum_i k_i^3 
\nonumber \\
&+& (-3\epsilon - \frac{3}{2}\eta -\frac{3}{2}s)
\sum_{i\neq j} k_i k_j^2
\nonumber \\
&+& ( 3\epsilon + \frac{3}{2}\eta +(\frac{5}{2}-2\bc_1)s) 
k_1 k_2 k_3
\nonumber \\
&+& (6(1+2\bc_1)\epsilon - 3(1-2\bc_1) \eta + (5+6\bc_1)s) 
\frac{1}{K} \sum_{i>j}
k_i^2 k_j^2 
\nonumber \\
&+& ( -6\bc_1 \epsilon +3(1-\bc_1)\eta -(1+6\bc_1)s)
\frac{1}{K^2} \sum_{i\neq j} k_i^2 k_j^3
\nonumber \\
&-& (1+\bc_1)s \frac{1}{K} \sum_i k_i^4 
- \bc_1 s\frac{1}{K^2} \sum_{i\neq j} k_i k_j^4 
\Bigg)
\label{A34}
\eea
and
\bea
\Delta \CA \supset -\frac{1}{4} \left(\frac{1}{c_s^2}-1\right)_K
\left(\epsilon + \frac{\eta}{2} + \frac{s}{2} \right) 
(N_1 + {\rm sym}) ~,
\label{A35}
\eea
where 
\bea
N_1 \equiv \frac{k_1}{2} \sum_i k_i^2 ~{\rm Re}
\int_0^{\infty} dx_1 \frac{1}{x_1^2} e^{-ix_1\frac{k_2+k_3}{k_1}}
(-1 - i\frac{k_2+k_3}{k_1}x_1 + \frac{k_2 k_3}{k_1} x_1^2) h^*(x_1)
~.
\eea

\subsection{Final results}
\label{AppCorrFinal}
Collecting all the results in Sec.~\ref{SecCubic} and this appendix,
we get the final result:
\begin{eqnarray}
\langle \zeta(\textbf{k}_1)\zeta(\textbf{k}_2)\zeta(\textbf{k}_3)\rangle
&=&
(2\pi)^7\delta^3(\textbf{k}_1+\textbf{k}_2+\textbf{k}_3)
(\tilde P_K^\zeta)^2
\frac{1}{\prod_i k_i^3} \cr &\times&
(\CA_\lambda +{\CA}_c + \CA_o +\CA_\epsilon +\CA_\eta +\CA_s)
\label{3pointFinalA}
\end{eqnarray}
where we have decomposed the shape of the three point function
into six parts
\begin{eqnarray}
\CA_\lambda &=& \left(\frac{1}{c_s^2}-1 
- \frac{2\lambda}{\Sigma} +(3-2\bc_1)l \frac{\lambda}{\Sigma} \right)_K
\frac{3k_1^2k_2^2k_3^2}{2K^3} ~, 
\label{AlamA} \\
\CA_c &=&
\left(\frac{1}{c_s^2}-1\right)_K
\left(-\frac{1}{K}\sum_{i>j}k_i^2k_j^2+\frac{1}{2K^2}
\sum_{i\neq j}k_i^2k_j^3+\frac{1}{8}\sum_{i}k_i^3 \right) ~, 
\label{AcA} \\
\CA_o &=& 
\left(\frac{1}{c_s^2}-1 - \frac{2\lambda}{\Sigma} \right)_K
\left(\epsilon F_{\lambda\epsilon} + \eta F_{\lambda\eta} + s F_{\lambda s}
 \right)
\nonumber \\
&+& \left(\frac{1}{c_s^2}-1 \right)_K
\left(\epsilon F_{c\epsilon} + \eta F_{c\eta} + s F_{cs} \right)
\label{AoA} \\
\CA_\epsilon &=& \epsilon \left( -\frac{1}{8}
\sum_i k_i^3 + \frac{1}{8}
\sum_{i\neq j} k_i k_j^2 + \frac{1}{K} \sum_{i>j} k_i^2 k_j^2
\right) ~, 
\label{AepA} \\ 
\CA_\eta &=& \eta \left( \frac{1}{8} \sum_i k_i^3
\right)~,
\label{AetaA} \\
\CA_s &=& s F_s ~.
\label{AsA}
\end{eqnarray}
The definitions of the sound speed $c_s$, $\Sigma$ and $\lambda$ are
\bea
c_s^2 &\equiv& \frac{P_{,X}}{P_{,X} + 2X P_{,XX}} ~,
\nonumber \\
\Sigma &\equiv& X P_{,X} + 2X^2 P_{,XX} ~,
\nonumber \\
\lambda &\equiv& X^2 P_{,XX} + \frac{2}{3} X^3 P_{,XXX} ~.
\eea
The definitions of the four slow variation parameters are
\bea
\epsilon \equiv -\frac{\dot H}{H^2} ~, ~~~~
\eta \equiv \frac{\dot \epsilon}{\epsilon H} ~, ~~~~
s \equiv \frac{\dot c_s}{c_s H} ~, ~~~~
l \equiv \frac{\dot \lambda}{\lambda H} ~.
\eea
$\tilde P^\zeta_K$ is defined as 
\bea
\tilde P^\zeta_K \equiv \frac{1}{8\pi^2} \frac{H_K^2}{c_{s K}\epsilon_K} ~.
\label{tPA}
\eea
Note that $H$, $c_s$, $\epsilon$, 
$\lambda$ and $\Sigma$ in this final result are 
evaluated at the moment $\tau_K \equiv -\frac{1}{Kc_{s K}} +\CO(\epsilon)$
when the wave number $K\equiv k_1 +k_2 +k_3$ exits the horizon
$K c_{s K} = a_K H_K$. So $\tilde P^\zeta_K$ in (\ref{tPA}) is defined 
differently from (\ref{Pk}).
The various functions $F$ are given by the following: 
\bea
F_{\lambda\epsilon} &\equiv& 
\left( \frac{3}{2}\bc_1 - \frac{9}{2}\bc_2 - \frac{39}{4}
- \frac{3}{2} \ln\frac{k_1k_2k_3}{K^3} \right)
\frac{k_1^2k_2^2k_3^2}{K^3}
\nonumber \\
&+& 
\frac{3}{4K^2}\sum_{i\neq j} k_i^2 k_j^3 - \frac{3}{2K} \sum_{i>j}
k_i^2 k_j^2  + \frac{3}{4} R(k_1,k_2,k_3) ~,
\label{Flambdaepsilon} \\
F_{\lambda\eta} &\equiv&
\half F_{\lambda\epsilon} -( 3\bc_1-\frac{33}{4})
\frac{k_1^2k_2^2k_3^2}{K^3} ~,
\label{Flambdaeta} \\
F_{\lambda s} &\equiv&
\half F_{\lambda\epsilon} +(\frac{3}{2}\bc_1 -6)\frac{k_1^2k_2^2k_3^2}{K^3} ~,
\label{Flambdas} \\
F_{c\epsilon} &\equiv& 
-\frac{1}{8}(\bc_1+3\bc_2+16) \sum_i k_i^3 - \frac{5}{4} \sum_{i\neq j} k_i 
k_j^2 + \frac{11}{8} k_1 k_2 k_3
\nonumber \\
&+& (\bc_1 + 3\bc_2 +\frac{9}{2})\frac{1}{K} 
\sum_{i>j} k_i^2 k_j^2
-(\frac{\bc_1}{2} +\frac{3}{2}\bc_2 + \frac{3}{2}) \frac{1}{K^2} 
\sum_{i\neq j} k_i^2 k_j^3
+ (\frac{9}{2}-3\bc_1) \frac{k_1^2 k_2^2 k_3^2}{K^3}
\nonumber \\
&+& \ln \frac{k_1 k_2 k_3}{K^3} 
\left(-\frac{1}{8} \sum_i k_i^3 + \frac{1}{K} \sum_{i>j} k_i^2 k_j^2 
-\frac{1}{2K^2} \sum_{i\neq j} k_i^2 k_j^3 \right) 
+ \frac{3}{4} Q(k_1,k_2,k_3) ~,
\label{Fcepsilon} \\
F_{c\eta} &\equiv&
-(\frac{\bc_1}{16} + \frac{3}{16}\bc_2 + \frac{9}{16}) \sum_i k_i^3
-\frac{11}{16} \sum_{i\neq j} k_i k_j^2
+\frac{11}{16} k_1 k_2 k_3
\nonumber \\
&+& (\frac{\bc_1}{2} + \frac{3\bc_2}{2} -\frac{1}{4}) \frac{1}{K}
\sum_{i>j} k_i^2 k_j^2 
+ (-\frac{\bc_1}{4} -\frac{3}{4}\bc_2 + \frac{1}{4}) \frac{1}{K^2} 
\sum_{i\neq j} k_i^2 k_j^3 
+(-\frac{9}{4} + \frac{3}{2}\bc_1) \frac{k_1^2 k_2^2 k_3^2}{K^3}
\nonumber \\
&+& \ln \frac{k_1 k_2 k_3}{K^3} 
\left(-\frac{1}{16} \sum_i k_i^3 + \frac{1}{2K} \sum_{i>j} k_i^2 k_j^2 
-\frac{1}{4K^2} \sum_{i\neq j} k_i^2 k_j^3 \right) 
+\frac{3}{8} Q(k_1,k_2,k_3) ~,
\label{Fceta} \\
F_{c s} &\equiv&
-(\frac{7\bc_1}{16} + \frac{3}{16}\bc_2 +\frac{21}{16}) \sum_i k_i^3
-\frac{5}{16} \sum_{i\neq j} k_i k_j^2
+(\frac{\bc_1}{4} + \frac{3}{16}) k_1 k_2 k_3
\nonumber \\
&+& (\frac{1}{2}\bc_1 + \frac{3}{2}\bc_2 +\frac{9}{4}) \frac{1}{K}
\sum_{i>j} k_i^2 k_j^2 
+ (\frac{19}{8}\bc_1 -\frac{3}{4}\bc_2 - \frac{7}{4}) \frac{1}{K^2} 
\sum_{i\neq j} k_i^2 k_j^3 
\nonumber \\
&+& \frac{1}{8}(1+\bc_1) 
\frac{1}{K} \sum_i k_i^4 
+ \frac{\bc_1}{8} \frac{1}{K^2} \sum_{i\neq j} k_i k_j^4
\nonumber \\
&+& (\frac{27}{4} - \frac{3}{2}\bc_1) \frac{k_1^2 k_2^2 k_3^2}{K^3}
+ (\frac{3}{4} - \frac{3}{2}\bc_1) \frac{1}{K^3} 
\sum_{i\neq j} k_i^2 k_j^4
+ (\frac{3}{2} -3 \bc_1) \frac{1}{K^3} 
\sum_{i>j} k_i^3 k_j^3
\nonumber \\
&+& \ln \frac{k_1 k_2 k_3}{K^3} 
\left(-\frac{1}{16} \sum_i k_i^3 + \frac{1}{2K} \sum_{i>j} k_i^2 k_j^2
-\frac{1}{4K^2} \sum_{i\neq j} k_i^2 k_j^3 \right)
+\frac{3}{8} Q(k_1,k_2,k_3) ~,
\label{Fcs} \\
F_s &\equiv&
-\frac{1}{4}\bc_1 \sum_i k_i^3 + \frac{1}{4} 
\sum_{i\neq j} k_i k_j^2 -\frac{1}{4} k_1 k_2 k_3
+(-2 +2\bc_1) \frac{1}{K} \sum_{i>j} k_i^2 k_j^2
\nonumber \\
&+& (\frac{3}{2}-\bc_1) \frac{1}{K^2} \sum_{i\neq j} k_i^2 k_j^3
+ (\frac{9}{2} -3\bc_1) \frac{k_1^2 k_2^2 k_3^2}{K^3} ~,
\eea
where $\bc_1 = 0.577 \cdots$ is the Euler constant and $\bc_2 \equiv
\bc_1-2+\ln 2 = -0.73 \cdots$. 
The functions $R(k_1,k_2,k_3)$ and $Q(k_1,k_2,k_3)$
involve special functions,
\bea
R(k_1,k_2,k_3) &\equiv& \frac{k_2^2 k_3^2}{k_1}
{\rm Re}\left[ \int_0^{\infty} dx_1~
\left( 1- i\frac{k_2+k_3}{k_1} x_1 \right)~
e^{-i\frac{k_2+k_3}{k_1}x_1}~ h^*(x_1) \right]
+ {\rm sym} ~,
\\
Q(k_1,k_2,k_3) &\equiv& 
- k_1 {\rm Re} \left[ \int_0^{\infty} dx_1 \frac{1}{x_1} 
\left(k_2^2 + k_3^2 +i k_2 k_3 \frac{k_2+k_3}{k_1} x_1 \right)~
e^{-i\frac{k_2+k_3}{k_1}x_1} ~\frac{dh^*(x_1)}{dx_1} \right]
\nonumber \\
&-& (k_2^3 + k_3^3) {\rm Re}
\left[ \int_0^{\infty} dx_K \frac{e^{-ix_K}}{x_K} \right]
\nonumber \\
&-& \frac{k_2^2 k_3^2}{k_1} {\rm Re} \left[
\int_0^\infty dx_1~ e^{-i\frac{k_2+k_3}{k_1}x_1} h^*(x_1) \right]
\nonumber \\
&-& \frac{k_1}{6} \sum_i k_i^2 ~{\rm Re} 
\Bigg[ \int_0^\infty \frac{dx_1}{x_1^2}
e^{-i\frac{k_2+k_3}{k_1}x_1} \left(-1-i\frac{k_2+k_3}{k_1}x_1
+ \frac{k_2 k_3}{k_1^2}x_1^2 \right) h^*(x_1) \Bigg]
\nonumber \\
&+& {\rm sym}~,
\label{Qexpress}
\\
h(x) &\equiv& - 2i e^{ix} + i e^{-ix} (1+ix) [\Ci(2x)+i\Si(2x)]
- i\pi \sin x + i\pi x \cos x ~,
\\
x_i &\equiv& -k_i c_{sK} \tau ~, ~~~~ x_K \equiv -K c_{sK} \tau ~.
\nonumber
\eea
In all formulae, the ``sym'' stands for two other terms with cyclic
permutation of the indices 1, 2 and 3.

\subsection{The squeezed limit}
\label{SecSqueezed}

It is interesting to look at the behaviors of various functions in the
squeezed limit (for example, $k_1=k_2$ and $k_3 \to 0$), 
because from them one can roughly know whether the
shape of a non-Gaussianity is closer to the DBI type
(Fig.~\ref{Fc}), or the slow-roll type (Fig.~\ref{Fep}). 

In slow-roll inflation, Maldacena has argued that the three-point
function in the squeezed limit goes to \cite{Maldacena:2002vr}
\bea
\langle \zeta(\bk_1) \zeta(\bk_2) \zeta(\bk_3) \rangle
\to - (2\pi)^7 \frac{1}{4k_1^3 k_3^3} (n_s-1) P^\zeta_{k_1}
P^\zeta_{k_3}
\label{Consistency}
\eea
to all orders in the slow-roll parameters. This condition was
generalized to general single-field inflation in
Ref.~\cite{Creminelli:2004yq}.

Here we check this condition in the general single-field inflation
model with large non-Gaussianities. 
For a large
$\lambda/\Sigma \gg 1$, the r.h.s.~of the consistency condition starts from
order $\CO(\epsilon)$ (does not include the terms $P^\zeta_{k_1}
P^\zeta_{k_3}$), while the l.h.s. starts from
$\CO(\lambda/\Sigma)$ in $\CA_{\lambda}$ and
$\CO(\epsilon\lambda/\Sigma)$ in $\CA_o$. So in order for the
condition to hold, both the leading and subleading order terms in the
three-point function have to vanish in the squeezed limit. It is not
difficult to see that $\CA_\lambda$ vanishes in this
limit. Interestingly, the
subleading terms (\ref{Flambdaepsilon}),
(\ref{Flambdaeta}) and (\ref{Flambdas}) also vanish in this limit due
to a cancellation from the special function $R$ (see Appendix
\ref{SecDetails}). 

The case with a small $c_s\ll 1$ is similar. The condition
(\ref{Consistency}) requires the leading and subleading
orders of the l.h.s.~vanish. It is easy to see the leading order
contribution $\CA_c$ satisfies this condition. 
For the subleading order $\CO(\epsilon/c_s^2)$,
(\ref{Fcepsilon}), (\ref{Fceta}) and (\ref{Fcs}) vanish
in this limit due to a cancellation from the special function $Q$ (see
Appendix \ref{SecDetails}).
Overall $\CA_o$ goes as
\bea
\frac{\CA_o}{k_1 k_2 k_3} \propto \frac{k_3}{k_1} ~.
\eea
$\CA_\epsilon$ and $\CA_\eta$ are the same as the slow-roll case;
in addition, since
\bea
\frac{F_s}{k_1 k_2 k_3} \to \frac{k_1}{4 k_3} ~,
\eea
the order $\CO(\epsilon)$ terms on both sides of the condition also
match, as we can see from (\ref{index1}).
So we have checked that all orders (the leading,
subleading and next-to-subleading orders) of our full results satisfy
the consistency condition.

\subsection{Some details}
\label{SecDetails}

In this section, we demonstrate some details that the first six 
$F$'s
vanish in the squeezed limit.
To calculate the squeezed limit of the $R$-term, we analytically
continue the integrand in the convergent direction by $x\to -ix$ and 
note the
asymptotic behavior
\begin{eqnarray}
\textrm{Ci}(-2ix)-i\textrm{Si}(-2ix)\sim
-\frac{i}{2}\pi-\frac{1}{2x}e^{-2x}(1+\mathcal{O}(\frac{1}{x})),~~~~x\rightarrow
+\infty ~.
\end{eqnarray}
So the $e^x$ terms cancel in the asymptotic behavior of
$h^*(x)$, and we find
\begin{eqnarray} \label{asymptoticinfinity}
h^*(-ix)\sim xe^{-x}, ~~~~x\rightarrow +\infty ~.
\end{eqnarray}
One can also compute the asymptotic of $h^*(x)$ near $x\sim
0$, and we find
\begin{eqnarray} \label{asymptoticzero}
h^*(x)=(-\bc_2 i-i\ln x)+\mathcal{O}(x^2) ~.
\end{eqnarray}
We can now take a squeezed limit $k_3\rightarrow 0$ and
$k_1=k_2 \equiv k$ in the $R$-term. 
There are 3 terms in the symmetric rotation of
the indices. Two terms are proportional to $k_3^2$ and the
integral is convergent because of the good asymptotic behavior
(\ref{asymptoticinfinity}), (\ref{asymptoticzero}), so they vanish
in the squeezed limit. The remaining term is
\begin{eqnarray}
R(k_1,k_2,k_3) &=& \frac{k^4}{k_3} {\rm Re}\left[-i
\int_0^{\infty} dx_3~ \left( 1- \frac{2k}{k_3} x_3 \right)~
e^{-\frac{2k}{k_3}x_3}~ h^*(x_3) \right] ~.
\end{eqnarray}
We only keep leading term in the squeezed
limit and drop higher powers of $k_3$. We can then change
integration variable $x_3\rightarrow \frac{k_3}{2k}y$ and expand
around $k_3=0$ using the formula (\ref{asymptoticzero}). We find
\begin{eqnarray}
R(k_1,k_2,k_3) &=& \frac{k^4}{2k} {\rm Re}\left[-i
\int_0^{\infty} dy~ \left( 1-  y \right)~ e^{-y}~
(c-i\ln(\frac{k_3}{2k}y))\right]+\mathcal{O}(k_3^2) 
\nonumber \\
&=& 
\frac{k^3}{2}+\mathcal{O}(k_3^2) ~.
\end{eqnarray}
So the squeezed limit of the term $F_{\lambda\epsilon}$
vanishes
\begin{eqnarray}
F_{\lambda\epsilon} 
=(\frac{3}{8}-\frac{3}{4}+\frac{3}{8})k^3+\mathcal{O}(k_3^2)
=\mathcal{O}(k_3^2)
~.
\label{FleSqueezed}
\end{eqnarray}

The $Q$-term is more complicated. In the same limit,
\bea
Q &\to& 2k^3 {\rm Re} \int_0^\infty
\frac{dy}{y^2} \left( -2y e^{-i\frac{y}{2}} 
\frac{d}{dy} h^*(\frac{y}{2})
+ (2i-3y)e^{-iy} \right)
\nonumber \\
&+& \frac{2k^3}{3} {\rm Re} \int_0^\infty \frac{dx_1}{x_1^2}e^{-ix_1} (1+ix_1)
h^*(x_1) \nonumber \\
&-& \frac{2}{3} k^3 ~,
\label{Qinter}
\eea
where the 1st line comes from the first and second lines of
(\ref{Qexpress}) and their two cyclic permutations, 
the 2nd line comes from the last line of (\ref{Qexpress}) and it
cyclic permutation $k_1\to k_2$, the 3rd line comes from the rest.
Similar to the previous case,
we have used the change of variable $x_i\to \frac{k_i}{2k}y$ and the
expansion of the function $h^*(\frac{k_3}{2k}y)$ around zero.
To further evaluate (\ref{Qinter}), we need the full expressions
\bea
h^*(x) &=& 2ie^{-ix} - \frac{\pi}{2} (1+ix)e^{-ix} 
-ie^{ix} (1-ix) {\rm Ei}(-2ix) ~, \\
\frac{dh^*}{dx} &=& (1-\frac{\pi}{2} x - \frac{i}{x}) e^{-ix}
- ix e^{ix} {\rm Ei}(-2ix) ~.
\eea
Plugging in these expressions, and making use of the integrals
\bea
&& {\rm Re}~ i\int_0^\infty \frac{dx}{x^2} (1+ix)e^{-ix} = 1 ~,
\label{integral1} \\
&& -i \int_0^\infty dx ~{\rm Ei}(-2ix) =\half ~,
\\
&& \int_0^\infty \frac{dx}{x^2} \left[ (-\frac{\pi}{2}-2x) \cos 2x 
+ (2-\pi x) \sin 2x - {\rm si}(2x) \right] = 0 ~,
\eea
we get
\bea
Q \to \frac{20}{3} k^3 ~.
\eea
So in the squeezed limit,
\bea
F_{c\epsilon} \to 2 F_{c\eta} \to 2 F_{cs} \to -5k^3 + \frac{3}{4} Q
\to 0 ~.
\eea
Although not explicitly demonstrated, one expects that they vanish as
$\CO(k_3^2)$ as we see explicitly in (\ref{FleSqueezed}), because in this
limit the $\CO(k_3)$ term vanishes due to isotropy \cite{Creminelli:2004yq}.

\end{document}